\documentclass[journal,10pt]{IEEEtran}

\usepackage{amssymb}
\usepackage{amsmath}
\allowdisplaybreaks[4]
\usepackage{cite}
\usepackage{url}
\usepackage{xcolor}
\usepackage{cite,graphicx,amsmath,amssymb}
\usepackage{subfigure}
\usepackage{citesort}
\usepackage{fancyhdr}
\usepackage{mdwmath}
\usepackage{mdwtab}
\usepackage{caption}
\usepackage{amsthm}

\newtheorem{remark}{Remark}
\newtheorem{theorem}{Theorem}

\newtheorem{lemma}{Lemma}

\newtheorem{corollary}{Corollary}

\makeatletter
\def\ScaleIfNeeded{%
\ifdim\Gin@nat@width>\linewidth \linewidth \else \Gin@nat@width
\fi } \makeatother

\begin{document}

\title{Cache-enabled HetNets with Limited Backhaul: A Stochastic Geometry Model}


\author{


Congshan~Fan,~\IEEEmembership{Student Member,~IEEE,}
        Tiankui~Zhang,~\IEEEmembership{Senior Member,~IEEE,} \\
        Yuanwei~Liu,~\IEEEmembership{Member,~IEEE }
        and Zhiming~Zeng,~\IEEEmembership{Member,~IEEE}

\thanks{C. Fan, T. Zhang and Z. Zeng are with Beijing University of Posts and Telecommunications, Beijing, (email:\{fcs, zhangtiankui, zengzm\}@bupt.edu.cn).}
\thanks{ Y. Liu is with Queen Mary University of London, London, (e-mail: yuanwei.liu@qmul.ac.uk).}
\thanks{
This work is supported by the National Natural Science Foundation of China (No. 61971060). }
}

\maketitle
\begin{abstract}
With the rapid explosion of data volume from mobile networks, edge caching has received significant attentions as an efficient approach to boost content delivery efficiency by bringing contents near users. In this article, cache-enabled heterogeneous networks (HetNets) considering the limited backhaul is analyzed with the aid of the stochastic geometry approach. A hybrid caching policy, in which the most popular contents are cached in the macro BSs tier with the deterministic caching strategy and the less popular contents are cached in the helpers tier with the probabilistic caching strategy, is proposed. Correspondingly, the content-centric association strategy is designed based on the comprehensive state of the access link, the cache and the backhaul link. Under the hybrid caching policy, new analytical results for successful content delivery probability, average successful delivery rate and energy efficiency are derived in the general scenario, the interference-limited scenario and the mean load scenario. The simulation results show that the proposed caching policy outperforms the most popular caching policy in HetNets with the limited backhaul. The performance gain is dramatically improved when the content popularity is less skewed, the cache capacity is sufficient and the helper density is relatively large. Furthermore, it is confirmed that there exists an optimal helper density to maximize the energy efficiency of the cache-enabled HetNets.
\end{abstract}

\begin{IEEEkeywords}
{E}dge caching, heterogeneous networks, limited backhaul, stochastic geometry
\end{IEEEkeywords}

\section{Introduction}
Heterogeneous networks (HetNets), where small base stations (SBSs) are embed into the existing macro cells, are widely deployed as an effective method to increase the data rate of the radio links. However exiting backhaul links fail to provide large capacity at a relatively affordable cost. The heavy traffic cause severe congestion in the backhaul link, making it a bottleneck in improving the system throughput. Numerous of research contributions reveal that the vast majority of the content requests are generated by duplicating downloads of a few popular contents~\cite{Amaldi2008}.
By taking advantage of the redundance of the content requests and the abundance of the cache resource, edge caching where content is cached in the base stations~\cite{6933434,8288090} or user equipments~\cite{7544526,7742334} has been proposed for backhaul traffic releasing and content access delay reducing~\cite{8387202}.
In cache-enabled HetNets, contents are proactively stored at BSs during off-peak time and users can get the content from the local cache. By densely deploying the SBSs and bringing the content closer to users, the cache-enabled HetNets can greatly improve the system performance. Meanwhile, the combination of the HetNets and the BSs caching make the access protocol, user association, resource allocation, caching strategy and content delivery great change.

For an in-depth sight into the cache-enabled HetNets, extensive contributions have been carried out. Femto-caching system was proposed in~\cite{6600983} for the first time. The author analyzed the expected downloading time under two types of the coded and the uncoded content placement and solved the optimum content assignment problem. The potential of the energy efficiency (EE) in the cache-enabled wireless access networks was explored in~\cite{7445129}. The condition when the EE can benefit from caching, the relationship between the EE and the memory and the maximal EE gain were analyzed sequentially. By effectively exploiting the multicast opportunities, ~\cite{6952688} designed a novel caching scheme to significantly reduce the traffic. However, the above research are conducted based on the regular hexagonal or the grid topology. The regular hexagonal topology and the grid topology are idealistic compared with the actual scenario and not suitable for modeling the large-scale networks. Stochastic geometry is an effective method to capture the randomness and the complexity of node distributions in the HetNets. In stochastic geometry models, the nodes are typically distributed according to the Poisson point processes (PPPs) in the two dimensional plane, enabling to characterize various system performance analytically, such as coverage probability, average rate, delay and so on.
\vspace{-0.15in}
\subsection{Related Works}
The cache-based content delivery in the three-tier HetNets was proposed in~\cite{7194828}, where BSs, relays and users cooperated to transmit contents. The outage probability and the average ergodic rate were theoretically elaborated. The energy efficiency of a cache enabled two-tier HetNet was analyzed in ~\cite{7882725} under co-channel and orthogonal channel deployment scenarios. The authors of ~\cite{7536893} investigated the average delay of users based on three different content popularity models.

The content placement in the aforementioned work is according to the deterministic caching strategy. An alternative strategy is the probabilistic caching strategy where a particular content is stored in the node with a caching probability.
The authors in~\cite{8288090} derived the closed-form expressions for the coverage probability and local delay experienced by a typical user with the probabilistic caching strategy.
The work in ~\cite{7435255} focused on the interplay of caching and spectrum sharing under the probabilistic caching framework in the HetNets and characterized the outage probability in serving users' requests over the coverage area. A hybrid cache-enabled HetNet where macro BSs with the traditional sub-6 GHz are overlaid by dense millimeter wave pico BSs was considered in~\cite{8401954}. The success probability and the area spectral efficiency were discussed under two user association strategies, namely, the maximum received power scheme and the maximum rate scheme. With the successful transmission probability as the performance metric, ~\cite{8377141} focused on the analysis of joint random caching and random DTX under two scenarios of the high mobility scenario and the static scenario. In addition to the static caching strategy, the dynamics of the cache replacement algorithm is incorporated into the analysis. An information centric modeling of the cache enabled cellular networks was investigated in ~\cite{7247159} where both the MBS and the SBS can perform caching and operate under the least recently used (LRU) based content eviction policy. Based on the obtained performance indicators, caching policy can be optimized accordingly.
The work in~\cite{7842078} investigated the successful offloading probability in the cache-enabled HetNets and obtained the optimal caching probability by maximizing the successful offloading probability.
In~\cite{7875124}, the author investigated the optimal caching policy by maximizing the success probability and the area spectral efficiency respectively and analyzed the impact of system settings.
Probabilistic content placement was studied in~\cite{7502130} to control cache-based channel selection diversity and network interference in a wireless caching helper network.
The research was extended to the most general N-tier HetNets in~\cite{8125744}, in which probabilistic tier-level content placement was analyzed given the network performance metric of the successful delivery probability. Different from the uncoded caching, coded caching can exploit the accumulated cache size by caching different segments of a content in different nodes. The coded caching in a large-scale small-cell network (SCN) was investigated in~\cite{7880694} and the performance was characterized by two metrics of the average fractional offloaded traffic and the average ergodic rate. The author of~\cite{7932139} proposed a combined coded caching strategy in disjoint cluster-centric SCNs and analyzed the successful content delivery probability.
\vspace{-0.15in}
\subsection{Motivation and Contributions}
The above research neglected the impact of backhaul link on the performance analysis in the cache-enabled cellular network. In the practical scenario, BSs can not store all contents due to the finite cache capacity. The uncached content should be retrieved from the core network via the backhaul link and delivered to users via the wireless link. It is necessary to analyze the performance of the cache-enabled HetNets by taking the backhaul link into account.

The backhaul link with the limited capacity makes constraint on the content delivery and introduces extra power consumption. In cache-enabled HetNets, the design of the caching policy, the access strategy and the backhaul assignment jointly determines the mode of the content delivery. The delivery rate for different delivery modes is affected by different factors. Unlike the wireless delivery rate is determined by the channel conditions, the backhaul delivery rate depends on the allocation of the limited backhaul capacity, and is generally smaller than the wireless delivery rate, which thus limits the overall content delivery rate.  Moreover, the backhaul power consumption is related to the backhaul delivery rate. In~\cite{7510749,7843834,7041201,7374694}, the authors conduct the backhaul-aware performance analysis of the cache-enabled cellular network based on a simplified backhaul model in which the backhaul delivery rate is set as the backhaul capacity.
The work of~\cite{8227679,7541306} studied the average delivery rate, in which the backhaul rate was obtained by the average splitting of backhaul capacity among users.
In~\cite{8904442}, the authors analyzed the cache performance in terms of successful content delivery probability (SCDP) taking backhaul capacity assigment based on the delivery rate demand into account.
By contrast, we focus on the content caching and corresponding content delivery in HetNets, which dramatically affect the cache performance.
Heterogeneous BSs differ in cache capacity and the density of nodes. Joint caching policies are required to be designed to take full advantages of the different cache ability of heterogeneous BSs to meet the content requests with different content popularity. Moreover, there exist different link states for heterogeneous BSs equipped with or without backhaul in the case of cache hit and miss. 
The association strategy should be coordinately designed by taking the caching policy and backhaul into account to make sure that the requested content is delivered with the proper link of the heterogeneous BSs.
To this end, we expand the research of the cache-enabled one-tier cellular network with the limited backhaul link in~\cite{fan2018energy}. We consider a two-tier HetNet in which the macro base stations (MBSs) tier is overlaid with the helpers tier. Both MBSs and helpers are equipped with caches while only the MBSs can connect to the core network via the backhaul link, helpers have no backhaul link. We analyze the system performance in cache-enabled HetNets with the limited backhaul. The main contributions are summarized as follows:

\begin{itemize}
  \item We propose a hybrid cache policy in which the most popular contents are cached in the MBSs tier with the deterministic caching strategy and the less popular contents are cached in the helpers tier with the probabilistic caching strategy. We design a corresponding association scheme by taking the overall consideration of the wireless channel condition, cache policy and the backhaul assignment. Based on the proposed network framework, we analyze the system performance with the aid of the stochastic geometry approach.
  \item We derive the tractable expression of the successful content delivery probability (SCDP) for three scenarios, specifically the general scenario, the interference-limited scenario and the mean load scenario. These expressions reveal that the SCDP is determined by three types of the network parameters which is respectively related to the access link, BSs caching and the backhaul link.
  \item We deduce the expression of the average successful delivery rate. The average successful delivery rate is verified to consist of two parts: the basic rate demand and the average of the extra rate exceeding the rate demand. With the results of the average successful delivery rate, we obtain the expression of the energy efficiency.
  \item Numerical results demonstrate that 1) the hybrid cache policy is able to improve the system performance effectively compared with the most popular caching strategy; 2) the performance gain is more obvious under the condition that the content popularity is less skewed, the cache capacity is sufficient and the helpers density is relatively large; 3) for a fixed cache capacity, there exists an optimal BS density ratio to maximize the energy efficiency.
\end{itemize}

The rest of the paper is organized as follows. Section II gives the system model for cache-enabled HetNet with the limited backhaul. In Section III, new analytical expressions for the successful content delivery probability, average successful delivery rate and energy efficiency are derived. Numerical results are presented in Section IV, which is followed by the conclusions in Section V.
\section{System Model}
The system model, including cellular network model, cache model and association model, is described in this section.
\subsection{Cellular Network Model}
\begin{figure} [t!]
\centering
\includegraphics[width= 3in, height=1.5in]{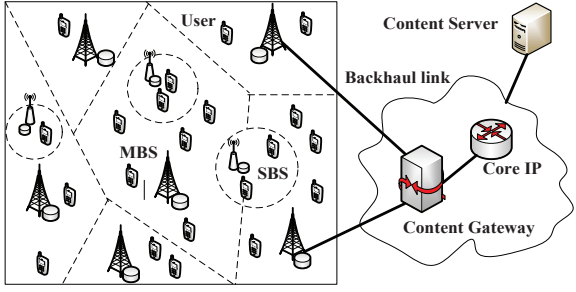}
 \caption{An example of cache-enabled HetNets structure.
  }
 \label{system_model}
\end{figure}

A two-tier cache-enabled HetNet is considered in this paper, where a macro base stations (MBSs) tier is overlaid with a dense helpers tier. Denoting ${x_{i,j}}$ as the location of the $j$ th BS in the $i$ th tier, the spatial distribution of the BSs in the $i$ th tier ${{\rm{\Phi }}_i} = \left\{ {{x_{i,j}},j = 0,1,2 \ldots } \right\}$ obeys independent Poisson Point Process (PPP) in the two dimensional Euclidean plane, and the intensity is ${\lambda _i}$. The locations of users (UEs) are also spatially distributed according to an independent PPP ${{\rm{\Phi }}_u}$ and the intensity is ${\lambda _u}$. MBSs are equipped with caches and connected to the core network via the backhaul link with limited capacity. Helpers have caches but no backhaul link, contents can be prefetched during off-peak times by broadcasting~\cite{6495773}. Fig.~\ref{system_model} shows an illustration of the network topology. According to the Palm theory, the statistical properties of UE at any position coincide with that of a typical UE at a fixed position~\cite{chiu2013stochastic}. Without loss of generality, the analysis is conducted on a typical UE at the origin, namely the tagged UE. $k \in \left\{ {1,2} \right\}$ is denoted as the index of the tier that the tagged UE is associated with.
\subsection{Channel Model}
The wireless channel gain consists of two types of propagation effect, i.e., path loss and Rayleigh fading. Denoting ${z_{i,j}}$ as the distance of the tagged UE from BS ${x_{i,j}}$ in the $i$ th tier, the path loss is calculated with the widely used power-law model as $z_{i,j}^{ - \alpha }$, where $2 < \alpha  \le 4$ is the path loss exponent. Rayleigh fading follows the independent and identically distributed (i.i.d.) exponential distribution with mean 1, ${h_{i,j}}\sim \exp \left(1 \right)$. Setting the transmission power of the BSs as ${P_i}$, the receive power at the tagged UE can be expressed as ${P_i}{h_{i,j}}z_{i,j}^{ - \alpha }$.
\subsection{Cache Model}
The content library is denoted as ${\rm{{\cal F}}} = \left\{ {{f_1},{f_2} \cdots {f_{N - 1}},{f_N}} \right\}$ and contains $N = \left| {\rm{{\cal F}}} \right|$ contents. All contents are assumed to have the same size $F$. For the scenarios of different content sizes, the same analysis is still applicable by splitting the content into chunks of equal size~\cite{7932139}. A large number of statistical results show that content popularity distribution changes slowly over time and can be approximated as static~\cite{6195469}. In this paper, the popularity of the content library follows Zipf distribution. Sorting the content in the descending order of popularity, the popularity of the $n$-ranked content is written as
{\setlength\abovedisplayskip{1.5pt}
\setlength\belowdisplayskip{1.5pt}
\begin{align}\label{Content Popularity}
{q_n} = \frac{{{n^{ - \gamma }}}}{{\sum\limits_{k = 1}^N {{k^{ - \gamma }}} }},
\end{align}}
where $\gamma  \ge 0$ is the shape parameter, reflecting the skewness of the popularity distribution. The larger the value of $\gamma$ is, the more uneven the content popularity distribution.

In HetNets,  MBSs have large transmission power and can ensure wide coverage range. Correspondingly, helpers are densely deployed to boost the network capacity. In order to ensure high cache hit ratio and increase content diversity at the same time, we propose a hybrid caching policy in this paper. Contents library is divided into two non-overlapping groups: the first group ${{\rm{{\cal F}}}_{Mp}} = \left\{ {{f_1},{f_2} \cdots {f_{{N_{Mp}} - 1}},{f_{{N_{Mp}}}}} \right\}$ contains the most popular ${N_{Mp}} = \left| {{{\rm{{\cal F}}}_{Mp}}} \right|$ contents, the second group ${{\rm{{\cal F}}}_{Lp}} = \left\{ {{f_{{N_{MP}} + 1}},{f_{{N_{Mp}} + 2}} \cdots {f_{N - 1}},{f_N}} \right\}$ contains the remaining less popular ${N_{Lp}} = \left| {{{\rm{{\cal F}}}_{Lp}}} \right|$ contents. MBSs and helpers are equipped with cache of different size to store ${N_1}$  and ${N_2}$ contents respectively. MBSs tier employs the deterministic caching strategy in which all MBSs store the same first contents group ${{\rm{{\cal F}}}_{Mp}}$ and ${N_1} = {N_{Mp}}$. As such, the most popular contents requested frequently by major users can obtain high cache hit ratio. Probabilistic caching strategy is performed in helpers tier. Each helper independently selects ${N_2}$ contents from the second content group ${{\rm{{\cal F}}}_{Lp}}$  to store in a random way and the caching probabilities of the contents in ${{\rm{{\cal F}}}_{Lp}}$ are the same, equal to ${p_{Lp}} = \frac{{{N_2}}}{{{N_{Lp}}}}$. Since the helpers density is comparatively large and each UE may be served by multiple helpers, probabilistic caching can make effective use of spatial content diversity to increase UEs' chance of obtaining the huge less popular contents.

The probability that the user requests the content from the first group is calculated as
{
\begin{align}\label{Content Request1}
{Q_{Mp}} = \sum\limits_{n = 1}^{{N_{MP}}} {{q_n}}.
\end{align}}

The probability that the user requests the content from the second group is calculated as
{\setlength\abovedisplayskip{1.5pt}
\setlength\belowdisplayskip{-10.5pt}
\begin{align}\label{Content Request2}
{Q_{Lp}} = 1 - {Q_{Mp}} = \sum\limits_{n = {N_{Mp}} + 1}^N {{q_n}}.
\end{align}}
\subsection{Cell Association}
Based on the hybrid caching policy, we adopt a new content-centric association strategy, where the user is associated with the strongest BS capable of accessing the requested content in the ways of local cache or backhaul link. For different content groups, the association strategies are specified as follows:

\vspace{0.1in}
\noindent\emph{1) Association Strategy of First Content Group}

In case the requested content belongs to the first group, user connects to the MBSs tier since the content is only cached in MBSs. In the MBSs tier, all MBSs store the same content group, then user is associated with the MBS with the maximum received power,
{
\begin{align}\label{Associate scheme1}
{x_0} = \arg \mathop {max}\limits_{{x_{1,j}} \in {\Phi _1}} {\rm{ }}{P_1}z_{1,j}^{ - \alpha }.
\end{align}}

In cache-enabled HetNets, MBSs have large transmission power and sufficient cache capacity, the combination of the maximum received power association and the deterministic caching strategy on the one hand greatly increases the delivery rate, and on the other hand improves the cache hit ratio.

Universal frequency reuse is adopted to improving the spectrum efficiency. When the user requests content from the first group, user is associated with the closet MBS. The total interference consists of all MBSs except the serving MBS and all helpers. The signal to interference plus noise ratio (SINR) is specified as:
\begin{align}\label{Associate scheme2}
{\rm SIN{R_n}} = \frac{{{P_1}{h}z^{ - \alpha }}}{{{I_1} + {I_2} + {\sigma ^2}}},{\rm{   }}{f_n} \in {{\cal F}_{Mp}},
\end{align}\
where ${I_1} = \sum\limits_{{x_{1,j}} \in {\Phi _1}\backslash {x_{0}}} {{P_1}{h_{1,j}}z_{1,j}^{ - \alpha }}$, ${I_2} = \sum\limits_{j \in {\Phi _2}} {{P_2}{h_{2,j}}z_{2,j}^{ - \alpha }}$ denote the interference from  MBSs tier, helpers tier respectively. ${\sigma ^2}$ is the additive white Gaussian noise (AWGN).

\vspace{0.1in}
\noindent\emph{2) Association Strategy of Second Content Group}

In case the requested content belongs to the second group, user can connect to either MBSs tier or helpers tier. In MBSs tier, no MBSs store the second contents group and the user needs to retrieve the requested content from the core network through the backhaul link. In the helpers tier, each helper independently selects contents to store according to the specific caching probability and the user can promptly get the requested content from the local cache. According to the properties of PPP, the distribution of the helpers caching and not caching the content ${f_n}$ can be modeled as the thinning PPP ${\Phi _{{n^ + },2}}$, ${\Phi _{{n^ - },2}}$, the density are ${p_{Lp}}{\lambda _2}$, $\left( {1 - {p_{Lp}}} \right){\lambda _2}$. Considering channel characteristics, BSs caching and backhaul link comprehensively, the user is assumed to be associated with the BS which has the maximum received power and is capable of accessing the requested content. The associated BS can be expressed as
{
\begin{align}\label{Associate scheme2}
{x_0} = \arg \mathop {max}\limits_{{x_{i,j}} \in {\Phi _1} \cup {\Phi _{{n^ + },2}}} {P_i}z_{i,j}^{ - \alpha }.
\end{align}}

Due to the ultra dense deployment of helpers and the overall low popularity of the second content group, the probabilistic caching and the corresponding association strategy can guarantee users' Quality of Service (QoS) and improve the cache hit ratio at the same time.

 When the user requests content from the second group, user connects to either the MBSs tier or the helpers tier. The serving BS may not be the closet BS and the interference can be divided into two types according to the content availability. The SINR of the user connecting to $k$ th tier is specified as:
\begin{align}\label{Associate scheme2}
{\rm SIN{R_n}} = \frac{{{P_k}{h_{k}}z_{k}^{ - \alpha }}}{{{I_{{n^ + }}} + {I_{{n^ - }}} + {\sigma ^2}}},{f_n} \in {{\cal F}_{Lp}},
\end{align}
where ${I_{{n^ + }}} = \sum\limits_{{x_{i,j}} \in {\Phi _1} \cup {\Phi _{{n^ + },2}}\backslash {x_{0}}} {{P_i}{h_{i,j}}z_{i,j}^{ - \alpha }}$, ${I_{{n^ - }}} = \sum\limits_{{x_{2,j}} \in {\Phi _{{n^ - },2}}} {{P_2}{h_{2,j}}z_{2,j}^{ - \alpha }}$ denote the interference from BSs capable of accessing the requested content and  BSs not capable of accessing the requested content respectively.
\section{Performance Analysis}
This section analyzes the successful content delivery probability, average successful delivery rate and energy efficiency of the cache-enabled Hetnets, which effectively quantify the QoS of users and the performance of the whole network.
\subsection{Successful Content Delivery Probability}
Successful content delivery probability (SCDP) is defined as the probability that the requested content of the tagged user is successful accessed and the delivery rate $R$ exceeds the rate demand ${R_0}$. SCDP helps to measure users' satisfaction with the achievable delivery rate with respect to the rate demand, and can be regarded as a metric of the QoS. Combining the content popularity ${q_n}$, the SCDP is given as
\begin{align}\label{SCDP def}
C = \sum\limits_{n = 1}^N {{q_n}} {\mathbb{P}}\left( {\frac{W}{L}{{\log }_2}(1 + {\rm SIN{R_n}}) > {R_0}} \right),
\end{align}
where $W$ denotes the system bandwidth, $L$ is the total number of UEs served by the tagged BS, namely the load. BSs allocate equal spectrum resource to the associated UEs.

In order to analyze the follow-up system performance, we derive some auxiliary results in advance. Since the caching policy and the cell association strategy for the first content group and the second content group is separately set irrespective of the content index, the related statistical characteristics like serving distance, association probability, load distribution and the system performance containing the SCDP, average successful delivery rate and energy efficiency are the same for different contents in two groups. For notational brevity, $\left\{ {{\rm Mp}\left., {{\rm Lp}} \right\}} \right.$ are used to uniformly mark two groups. In addition, we define ${\hat \lambda _{j,k}} = \frac{{{\lambda _j}}}{{{\lambda _k}}}$, ${\hat P_{j,k}} = \frac{{{P_j}}}{{{P_k}}}$ as the BSs density ratio and the transmit power ratio.

\vspace{0.1in}
\noindent \emph{1) Association Probability}

 Users requesting content from the first group can only connect to the MBSs tier, the association probability is expressed as
{
\begin{align}\label{Associate P1}
{A_{Mp,1}} = 1.
\end{align}}

Correspondingly, users requesting content from the second group can connect to either the MBSs tier or the helpers tier, the association probability is derived in the following lemma.
\begin{lemma}\label{lemma:Associate P2}
The probability that the tagged UE requesting the content from the second group connects to the $k$ th BSs tier is given by
\vspace{-0.1in}
\begin{align}\label{Associate P2}
{A_{Lp,k}} = \frac{{{{\lambda '}_k}P_k^{2/\alpha }}}{{\sum\limits_{i = 1}^2 {{{\lambda '}_i}} P_i^{2/\alpha }}},
\end{align}
where ${\lambda '_1} = {\lambda _1}$,${\lambda '_2} = {p_{Lp}}{\lambda _2}$ denote the equivalent MBSs density and helpers density of the content-centric HetNets.
\begin{proof}
\emph{If user requests content from the second group, user can access to the content from either the backhaul link of all MBSs or the local cache of part helpers. As a result, the network capable of accessing to the requested content is equivalent to a content-centric HetNets in which the distribution of MBSs and helpers follow two thinning PPPs and the corresponding tier density are ${\lambda '_1} = {\lambda _1}$, ${\lambda '_2} = {p_{Lp}}{\lambda _2}$. Matching the content-centric HetNets with the traditional HetNets, the association probability is derived by straightforwardly modifying the Lemmas 1 in~\cite{6287527}.}
\end{proof}
\end{lemma}
\begin{remark}
The tagged user prefers to connect to the tier with higher equivalent BS density ${\lambda '_k}$ and transmit power ${P_k}$ in the content-centric HetNets.
\end{remark}
According to the content-centric association, users connecting to the MBSs tier fall into two categories: MBSs cache hit users and MBSs cache miss users, namely users who request content from the first group and users who request content from the second group and associate with MBSs. To sum up, the probability to connect to the MBSs tier is expressed as
\begin{align}\label{tier P1}
&{A_1} = \sum\limits_{n = 1}^{{N_1}} {{q_n}} {A_{Mp,1}} + \sum\limits_{n = {N_1} + 1}^N {{q_n}} {A_{Lp,1}} \nonumber\\
&= {Q_{Mp}} + {Q_{Lp}}\frac{{{\lambda _1}P_1^{2/\alpha }}}{{{\lambda _1}P_1^{2/\alpha } + {p_{Lp}}{\lambda _2}P_2^{2/\alpha }}}.
\end{align}

With the conditional probability formula, the probability that user connecting to the MBS needs to fetch the requested content from the backhaul link is given by
\begin{align}\label{Backhaul ratio}
{p_b} = \frac{{{Q_{Lp}}{A_{Lp,1}}}}{{{A_1}}}.
\end{align}

By comparison, users connecting to the SBSs tier is users requesting content from the second group and associating with the helpers, namely the helpers cache hit users. The probability to connect to the helpers tier is expressed as
{
\begin{align}\label{tier P2}
{A_2} = {Q_{Lp}}{A_{Lp,2}}.
\end{align}}

\vspace{0.05in}
\noindent\emph{2) Load Distribution}

The load is the number of users served by the tagged BS, and the tagged user is included. Based on the lemma 3 of~\cite{6497002}, the distribution of the load of in the $k$ th tier is expressed as
\begin{align}\label{Load distribution}
&{P_{{L_k}}}({L_k} = {l_k} + 1) \nonumber\\
&= \frac{{{{3.5}^{3.5}}}}{{{l_k}!}}\frac{{\Gamma ({l_k} + 4.5)}}{{\Gamma (3.5)}}{\left( {\frac{{{A_k}{\lambda _u}}}{{{\lambda _k}}}} \right)^{{l_k}}}{\left( {3.5 + \frac{{{A_k}{\lambda _u}}}{{{\lambda _k}}}} \right)^{ - ({l_k} + 4.5)}},
\end{align}
where ${A_k}$ is the association probability as \eqref{tier P1}, \eqref{tier P2}.

The mean load in the $k$ th tier is approximated as
{
\begin{align}\label{Mean Load}
{\bar L_k} \approx {\mathbb{E}}({L_k}) = 1 + 1.28\frac{{{A_k}{\lambda _u}}}{{{\lambda _k}}}.
\end{align}}

\vspace{0.05in}
\noindent\emph{3) Active Probability}

In dense Hetnets, the intensity of helpers is comparable to or even higher than the intensity of users, some helpers may have no users to serve. In order to mitigate the interference and save the power consumption, inactive helpers should be turned off. The rest helpers that have users to serve are referred as active helpers. The probability that a helper is active is derived as~\cite{6497002}.
${f_S}\left( x \right)$ is the PDF of the coverage area of the helpers, and can be approximated as
\begin{align}\label{Coverage area}
{f_S}\left( x \right) \approx \frac{{{{3.5}^{3.5}}}}{{\Gamma \left( {3.5} \right)}}{\left( {\frac{{{\lambda _2}}}{{{A_2}}}} \right)^{3.5}}{x^{2.5}}{e^{ - 3.5{\lambda _2}x/{A_2}}}.
\end{align}
Active probability can be derived as
\normalsize
\begin{align}
\begin{array}{l}
{p_a} = 1 - \int_0^\infty  {{e^{ - {\lambda _u}x}}{f_S}\left( x \right)dx}\\
 = 1 - \int_0^\infty  {\frac{{{{3.5}^{3.5}}}}{{\Gamma \left( {3.5} \right)}}{{\left( {\frac{{{\lambda _2}}}{{{A_2}}}} \right)}^{3.5}}{x^{2.5}}{e^{ - 3.5{\lambda _2}x/{A_2} - {\lambda _u}x}}dx}
\end{array},
\end{align}
\normalsize
Let $3.5{\lambda _2}x/{A_2} - {\lambda _u}x = t$,
\begin{align}\label{Active probability}
\begin{array}{l}
{p_a} = 1 - \int_0^\infty  {\frac{{{{3.5}^{3.5}}}}{{\Gamma \left( {3.5} \right)}}{{\left( {\frac{{{\lambda _2}}}{{{A_2}}}} \right)}^{3.5}}\frac{1}{{{{\left( {3.5{\lambda _2}/{A_2} - {\lambda _u}} \right)}^{3.5}}}}{t^{2.5}}{e^{ - t}}dt}\\
 = 1 - {\left( {1 + \frac{{{A_2}{\lambda _u}}}{{3.5{\lambda _2}}}} \right)^{ - 3.5}}
\end{array}.
\end{align}
\normalsize

Based on the combination of the caching policy and association strategy, the SCDP can be specifically divided into three cases. In the first case, the user requests content from the first group and connects to the MBSs tier. Since the requested content is stored in the local cache, the user can directly get the content from the serving MBS and the delivery rate is equal to the access delivery rate of the MBS; In the second case, the user requests content from the second group and connects to the helpers tier. Due to the probabilistic caching strategy, the user can get the content from the serving helper and the delivery rate is equal to the access delivery rate of the helper; In the third case, the user requests content from the second group and connects to the MBSs tier. Since the requested content is not stored in the local cache, the user need the serving MBS to retrieve the requested content from the core network via the backhaul link and forward it to the user via the access link. The delivery rate is composed of the access delivery rate and the backhaul delivery rate. To sum up, SCDP is expressed as
{

\begin{align}\label{SCDP calculation}
&\Pr (R > {R_0})\nonumber\\
 &= {Q_{Mp}}\Pr \left( {{R_{Mp,1}} \ge {R_0}} \right) + {Q_{Lp}}\Pr \left( {{R_{Lp,2}} \ge {R_0},{A_{Lp,2}}} \right)\nonumber\\
& + {Q_{Lp}}\Pr \left( {R_{Lp,1}^w \ge {R_0}\& R_{Lp,1}^b \ge {R_0},{A_{Lp,1}}} \right)\nonumber\\
&\mathop  = \limits^{\left( a \right)} {Q_{Mp}}\underbrace {\Pr \left( {{R_{Mp,1}} \ge {R_0}} \right)}_{{\rm SADP} {\ }{C_{Mp}} } + {Q_{Lp}}\underbrace {\Pr \left( {{R_{Lp,2}} \ge {R_0},{A_{Lp,2}}} \right)}_{{\rm SADP}{\ } {C_{Lp,2}}}\nonumber\\
& + {Q_{Lp}}\underbrace {\Pr \left( {R_{Lp,1}^w \ge {R_0},{A_{Lp,1}}} \right)}_{{\rm SADP}{\ } C_{_{Lp,1}}^w}\underbrace {\Pr \left( {R_{Lp,1}^b \ge {R_0},{A_{Lp,1}}} \right)}_{{\rm SBDP}{\ } C_{_{Lp,1}}^b}
\end{align}}
\normalsize
Equation (a) makes sense due to the independence of the access link and the backhaul link. It is shown in \eqref{SCDP calculation} that SCDP depends on four factors, namely the successful access delivery probability (SADP) ${C_{Mp}}$, SADP ${C_{Lp,2}}$, SADP $C_{Lp,1}^w$, the successful backhaul delivery probability (SBDP) $C_{Lp,1}^b$.

In the following, we analyze four probabilities in sequence. To this end, two functions are defined to facilitate the description of two types of interference coming from the BSs with and without the requested content respectively:
\begin{align}
& G\left( {x,y} \right) = \frac{{2x}}{{y - 2}}{}_2{F_1}\left( {1,1 - \frac{2}{y};2 - \frac{2}{y}; - x} \right), {\label{formula 1} }\\
& H\left( {x,y} \right) = \frac{2}{y}{x^{2/\alpha }}B\left( {\frac{2}{y},1 - \frac{2}{y}} \right),\label{formula 2}
\end{align}
where ${}_2{F_1}\left( \cdot \right)$ is the Gauss hypergeometric function, $B\left( \cdot \right)$ is the Beta function.

SADP ${C_{Mp}}$  is the probability that the access delivery rate of MBSs is higher than the rate demand when user requests content from the first group.
\begin{lemma}\label{lemma:SADP 1}
The successful access delivery probability ${C_{Mp}}$ is expressed as
{
\begin{align}\label{SADP 1}
&{C_{Mp}} =  \int_0^\infty  {\sum\limits_{{l_1} = 0}^\infty  {2\pi {\lambda _1}z\exp ( - {z^\alpha }{P_1^{ - 1}}{\delta _1}{\sigma ^2})}}\nonumber\\
& \times \exp \left( { - \pi {\lambda _1}{z^2}\left( { G\left( {{\delta _1},\alpha } \right) +{p_a}{{\hat \lambda }_{2,1}}\hat P_{2,1}^{2/\alpha }H\left( {{\delta _1},\alpha } \right)+1 } \right)} \right) \nonumber\\
& \times{P_{{L_1}}}({l_1} + 1)dz,
\end{align}}
\normalsize

where ${\delta _1} = {2^{{l_1}\frac{{{R_0}}}{W}}} - 1$.
\vspace{0.1in}
\begin{proof}
\emph{Please refer to Appendix A.}
\end{proof}
\end{lemma}
SADP ${C_{Lp,2}}$ is the probability that the access delivery rate of helpers is higher than the rate demand when user requests content from the second group and connects to the helpers tier.
\begin{lemma}\label{lemma:SADP 2}
The successful access delivery probability ${C_{Lp,2}}$ is expressed as
{
\begin{align}\label{SADP 2}
&{C_{Lp,2}} = \int_0^\infty  \sum\limits_{{l_2} = 0}^\infty  2\pi {p_{Lp}}{\lambda _2}z{\exp \left( { - {z^\alpha }P_2^{ - 1}{\delta _2}{\sigma ^2}} \right)}\nonumber\\
&\times \exp \left( { - \pi {p_{Lp}}{\lambda _2}{z^2}\left( {{\xi _2}G\left( {{\delta _2},\alpha } \right) + {\varsigma _2}H\left( {{\delta _2},\alpha } \right) + {\zeta _2}} \right)} \right)\nonumber\\
&\times{P_{{L_2}}}\left( {{l_2} + 1} \right)dz,
\end{align}}

where ${\delta _2} = {2^{{l_2}\frac{{{R_0}}}{W}}} - 1$, coefficients for $k$th tier are given as ${\xi _k} = \left( {{\lambda _1}{P_1} + {p_a}{p_{Lp}}{\lambda _2}{P_2}} \right) {\left({\lambda '_k}{P_k}\right)}^{ - 1}$, ${\varsigma _k} = \left( {{p_a}\left( {1 - {p_{Lp}}} \right){\lambda _2}{P_2}} \right){\left({\lambda '_k}{P_k}\right)}^{ - 1}$, ${\zeta _k} = \left( {{\lambda _1}{P_1} + {p_{Lp}}{\lambda _2}{P_2}} \right){\left({\lambda '_k}{P_k}\right)}^{ - 1}$

\vspace{0.1in}
\begin{proof}
\emph{Please refer to Appendix B.}
\end{proof}
\end{lemma}

SADP ${C_{Lp,1}}$  is the probability that the access delivery rate of MBSs is higher than the rate demand when user requests content from the second group and connects to the MBSs tier.
\begin{lemma}\label{lemma:SADP 3}
The successful access delivery probability ${C_{Lp,1}}$ is expressed as
\begin{align}\label{SADP 3}
&C_{Lp,1}^w = \int_0^\infty  \sum\limits_{{l_1} = 0}^\infty  2\pi {\lambda _1}z{\exp \left( { - {z^\alpha }P_1^{ - 1}{\delta _1}{\sigma ^2}} \right)}\nonumber\\
&\times \exp \left( { - \pi {\lambda _1}{z^2}\left( {{\xi _1}G\left( {{\delta _1},\alpha } \right) + {\varsigma _1}H\left( {{\delta _1},\alpha } \right) + {\zeta _1}} \right)} \right)\nonumber\\
&\times{P_{{L_1}}}\left( {{l_1} + 1} \right)dz,
\end{align}
\begin{proof}
\emph{Please refer to Appendix C.}
\end{proof}
\end{lemma}

SBDP $C_{Lp,1}^b$  is the probability that the backhaul delivery rate of MBSs is higher than the rate demand when user requests content from the second group and connects to the MBSs tier. $C_{Lp,1}^b$ is determined by the backhaul assignment among users. Since the backhaul capacity is limited, the maximum number of users supported to deliver the content at the required delivery rate is fixed and can be calculated as ${N_b} = \frac{{{C_b}}}{{{R_0}}}$. Supposing the number of users accessing the requested content via the backhaul link in the MBS is ${N_{miss
}}$, if ${N_{miss}} \le {N_b}$, all ${N_{miss}}$ users can be scheduled with the backhaul link, otherwise, the backhaul link fails to support ${N_{miss}}$  users, and will randomly picks ${N_{b}}$ users.

Assuming that the requested content of the tagged user needs to be fetched through the backhaul link and there are another $m$  MBSs cache miss users associated with the tagged MBS, the successful backhaul delivery probability is expressed as
\vspace{-0.1in}
\begin{align}\label{SBDP}
&C_{Lp,1}^b\left( {{l_1} + 1} \right) = \left( {R_{Lp,1}^b \ge {R_0}|{L_1} = {l_1} + 1} \right)\nonumber \\
& = \sum\limits_{m = 0}^{{N_b} - 1} {{{l}_1} \choose m} {\left( {1 - {p_b}} \right)^{{l_1} - m}}{\left( {{p_b}} \right)^m} \nonumber \\
&+ \sum\limits_{m = {N_b}}^{{l_1}} {{{l}_1} \choose m} {\left( {1 - {p_b}} \right)^{{l_1} - m}}{\left( {{p_b}} \right)^k}\frac{{{N_b}}}{{\left( {m + 1} \right)}}\nonumber \\
& = \sum\limits_{m = 0}^{{l_1}}{{{l}_1} \choose m} {\left( {1 - {p_b}} \right)^{{l_1} - m}}{\left( {{p_b}} \right)^k}\min \left\{ {1,\frac{{{N_b}}}{{\left( {m + 1} \right)}}} \right\}.
\end{align}
\normalsize
\begin{remark}
As the backhaul capacity tends to infinity, the number of the cache miss users that can be supported by the backhaul link ${N_b}$ is much greater than the actual number of cache miss users $m+1$ in Hetnets. $C_{Lp,1}=1$, the backhaul link has no effect on SCDP.
\end{remark}

Substituting \eqref{SADP 1}, \eqref{SADP 2}, \eqref{SADP 3} and \eqref{SBDP} in \eqref{SCDP calculation}, the SCDP is obtained.

\begin{remark}
SCDP is determined by three network parameters sets. The first set contains the access link related parameters: BSs density, transmit power and path loss parameter. The second set has correlation with the cache, including the content popularity and cache capacity. The third set is related to the backhaul link, namely the backhaul capacity.
\end{remark}
Due to the combination of various operations, such as the improper integral of the serving distance, the infinite summation over the load, the judgment of the minimum and the especial use of lookup tables for ${}_2{F_1}$, the SCDP expression for the general case is extremely complicated. In order to analyze the SCDP more conveniently, we derive two approximate results containing the interference-limited scenario and the mean load scenario in the following corollaries.

\begin{corollary}\label{corollary:SCDP NO noise}
The successful content delivery probability in the interference-limited scenario is given by
{
\begin{align}\label{SBDP NO noise}
&\tilde C = {Q_{Mp}}{{\tilde C}_{Mp}} + {Q_{Lp}}{{\tilde C}_{Lp,2}} + {Q_{Lp}}\tilde C_{Lp,1}^wC_{Lp,1}^b, \nonumber \\
&{{\tilde C}_{Mp}} = \sum\limits_{{l_1} = 0}^\infty  {{\left({G\left( {{\delta _1},\alpha } \right) + {p_2}{{\hat \lambda }_{2,1}}\hat P_{2,1}^{2/\alpha }H\left( {{\delta _1},\alpha } \right) + 1}\right)}^{-1}} \nonumber \\
&\times{P_{{L_1}}}({l_1} + 1), \nonumber \\
&{{\tilde C}_{Lp,2}} = \sum\limits_{{l_2} = 0}^\infty  {  {\left( {{\xi _2}G\left( {{\delta _2},\alpha } \right) + {\varsigma _2}H\left( {{\delta _2},\alpha } \right) + {\zeta _2}} \right)}^{-1}{P_{{L_2}}}\left( {{l_2} + 1} \right)},  \nonumber \\
&\tilde C_{Lp,1}^w = \sum\limits_{{l_1} = 0}^\infty  {  {\left( {{\xi _1}G\left( {{\delta _1},\alpha } \right) + {\varsigma _1}H\left( {{\delta _1},\alpha } \right) + {\zeta _1}} \right)}^{-1}{P_{{L_1}}}\left( {{l_1} + 1} \right)}.
\end{align}}
\normalsize

\vspace{0.1in}
\begin{proof}
\emph{In interference-limited scenario, the noise power is very small compared to the interference power. ${\tilde C_{Mp}}$, ${\tilde C_{Lp,2}}$ and $\tilde C_{Lp,1}^w$ can be obtained by letting the noise power tend to zero ${\sigma ^2} \to 0$ and calculating the integral of the serving distance in \eqref{SADP 1}, \eqref{SADP 2} and \eqref{SADP 3} respectively. The noise power has no effect on SBDP $C_{Lp,1}^b$. Substituting ${\tilde C_{Mp}}$, ${\tilde C_{Lp,2}}$ and $\tilde C_{Lp,1}^w$ and the unchanged $C_{Lp,1}^b$ into \eqref{SCDP calculation}, Corollary is proved.}
\end{proof}
\end{corollary}

\vspace{0.1in}
\begin{corollary}\label{corollary:SCDP mean load}
The successful content delivery probability with the mean load in the interference-limited scenario is given by
{
\begin{align}\label{SCDP mean load}
&\bar C = {Q_{Mp}}{{\bar C}_{Mp}} + {Q_{Lp}}{{\bar C}_{Lp,2}} + {Q_{Lp}}\bar C_{Lp,1}^w\bar C_{Lp,1}^b, \nonumber \\
&{{\tilde C}_{Mp}} = {{\left({G\left( {{{\bar \delta } _1},\alpha } \right) + {p_2}{{\hat \lambda }_{2,1}}\hat P_{2,1}^{2/\alpha }H\left( {{{\bar \delta } _1},\alpha } \right) + 1}\right)}^{-1}}, \nonumber \\
&{{\tilde C}_{Lp,2}} =  {\left( {{\xi _2}G\left( {{{\bar \delta } _2},\alpha } \right) + {\varsigma _2}H\left( {{{\bar \delta } _2},\alpha } \right) + {\zeta _2}} \right)}^{-1},  \nonumber \\
&\tilde C_{Lp,1}^w = {\left( {{\xi _1}G\left( {{{\bar \delta } _1},\alpha } \right) + {\varsigma _1}H\left( {{{\bar \delta } _1},\alpha } \right) + {\zeta _1}} \right)}^{-1},\nonumber \\
&\bar C_{Lp,1}^b = \sum\limits_{m = 0}^{{{\bar L}_1} - 1} {{{\bar L}_1-1} \choose m} {\left( {1 - {p_b}} \right)^{{{\bar L}_1} - m - 1}}{\left( {{p_b}} \right)^m}\nonumber \\
&\times \min \left\{ {1,\frac{{{N_b}}}{{\left( {m + 1} \right)}}} \right\},
\end{align}}
\normalsize
where ${\bar \delta _1} = {2^{{{\bar L}_1}\frac{{{R_0}}}{W}}} - 1$, ${\bar \delta _2} = {2^{\left( {{{\bar L}_2} + 1} \right)\frac{{{R_0}}}{W}}} - 1$, ${\bar L_1}$, ${\bar L_2}$ denote the mean load of the MBSs tier and the helpers tier.

\begin{proof}
\emph{${\bar C_{Mp}}$, ${\bar C_{Lp,2}}$, $\bar C_{Lp,1}^w$ and $\bar C_{Lp,1}^b$ can be obtained by calculating the mean load as \eqref{Mean Load} and eliminating the summation over the load ${l_k}$  in \eqref{SADP 1}, \eqref{SADP 2}, \eqref{SADP 3} and \eqref{SBDP} respectively. Using the approximation ${\mathbb{E}}{_L}\left( {C\left( l \right)} \right) \approx C\left( {{\mathbb{E}}{_L}} \right)$ and substituting ${\bar C_{Mp}}$, ${\bar C_{Lp,2}}$, $\bar C_{Lp,1}^w$ and $\bar C_{Lp,1}^b$ into (18), the corollary is proved.}
\end{proof}
\end{corollary}
\subsection{Average successful delivery rate}
Average successful delivery rate is defined as the average delivery rate of the tagged UE under the premise of the successful content delivery ${R_{suc}} = {\mathbb{E}}\left( {R|R \ge {R_0}} \right)$. Similar to the SCDP, the average successful delivery rate ${R_{suc}}$ has three cases.
\begin{itemize}
  \item If the user requests content from the first group, content can be obtained directly from the MBSs and ${R_{suc}}$ is equal to the average successful access delivery rate of the MBSs $R_{_{Mp,1}}^{suc}$, namely the average access delivery rate of MBSs on the condition that the access delivery rate exceeds the rate demand.
  \item If the user requests content from the second group and connects to the helpers tier, content can be obtained from the helpers and ${R_{suc}}$ is equal to the average successful access delivery rate of the helpers $R_{_{Lp,2}}^{suc}$, namely the average access delivery rate of helpers on the condition that the access delivery rate exceeds the rate demand.
  \item If the user requests content from the second group and connects to the MBSs tier, content needs to be retrieved from the core network through the backhaul link and ${R_{suc}}$ is equivalent to the average successful backhaul delivery rate $R_{_{Lp,1}}^{suc}$ set to the rate demand ${R_0}$ on the premise that access link and backhaul link are successfully provided.
\end{itemize}

According to the total probability law, ${R_{suc}}$ is expressed as
{
\begin{align}\label{Rate calculation}
{R^{suc}} &= {Q_{Mp}}R_{Mp,1}^{suc} + {Q_{Lp}}R_{Lp,2}^{suc} \nonumber \\
&+ {Q_{Lp}}{A_{Lp,1}}C_{Lp,1}^wC_{Lp,1}^bR_{Lp,1}^{suc}.
\end{align}}

$R_{_{Mp,1}}^{suc}$ and $R_{_{Lp,2}}^{suc}$ are given in the following theorems successively.

\begin{lemma}\label{lemma:Rate 1}
The average successful access delivery rate of MBSs $R_{_{Mp,1}}^{suc}$ is expressed as (29) at the top of
next page, where ${G^ * }\left( {x1,x2,y} \right) = G\left( {x1,y } \right) - G\left( {{x2},y} \right)$, ${H^ * }\left( {x1,x2,y} \right)  = H\left( {x1,y} \right) - H\left( {x2,y} \right)$, $\delta _1^r = {2^{\left( {{l_1} + 1} \right)\frac{r}{W}}} - 1$, ${\delta _1} = {2^{\left( {{l_1} + 1} \right)\frac{{{R_0}}}{W}}} - 1$.
\begin{figure*}[!t]
\normalsize
\begin{align}\label{Rate 1}
R_{Mp,1}^{suc} = {R_0} &+ \int_{{R_0}}^\infty  \int_0^\infty  \sum\limits_{{l_1} = 0}^\infty  2\pi {\lambda _1}z\exp \left( { - {{{z^\alpha }P_1^{-1}\left( {\delta _1^r - {\delta _1}} \right){\sigma ^2}}}}\right) \nonumber \\
& \times \exp \left( { - \pi {\lambda _1}{z^2}\left( {{G^ * }\left( {\delta _1^r,{\delta _1},\alpha } \right) + {{\hat \lambda }_{2,1}}\hat P_{2,1}^{2/\alpha }{H^ * }\left( {\delta _1^r,{\delta _1},\alpha } \right) + 1} \right)} \right) {P_{{L_1}}}({l_1} + 1)dzdr,
\end{align}
\hrulefill \vspace*{0pt}
\end{figure*}
\begin{proof}
\emph{Please refer to Appendix D.}
\end{proof}
\end{lemma}

\begin{lemma}\label{lemma:Rate 2}
The average successful access delivery rate of SBSs $R_{_{Lp,2}}^{suc}$ is expressed as expressed as (30) at the top of
next page, where $\delta _2^r = {2^{\left( {{l_2} + 1} \right)\frac{r}{W}}} - 1$, ${\delta _2} = {2^{\left( {{l_2} + 1} \right)\frac{{{R_0}}}{W}}} - 1$.
\begin{figure*}[!t]
\normalsize
{\begin{align}\label{Rate 2}
R_{Lp,2}^{suc}=& {A_{Lp,2}}{R_0} +  \int_{{R_0}}^\infty  {\int_0^\infty  {\sum\limits_{{l_2} = 0}^\infty  {2\pi {p_{Lp}}{\lambda _2}z\exp \left( { - {{{z^\alpha }P_2^{-1}\left( {\delta _2^r - {\delta _2}} \right){\sigma ^2}}}}\right)} } } \nonumber \\
&\times \exp \left( { - \pi {p_{Lp}}{\lambda _2}{z^2}\left( {{\xi _2}{G^ * }\left( {\delta _2^r,{\delta _2},\alpha } \right) + {\varsigma _2}{H^ * }\left( {\delta _2^r,{\delta _2},\alpha } \right) + {\zeta _2}} \right)} \right){P_{{L_2}}}\left( {{l_2} + 1} \right)dzdr,
\end{align}}
\hrulefill \vspace*{0pt}
\end{figure*}
\begin{proof}
\emph{Proof is similar to Appendix D.}
\end{proof}
\end{lemma}

Plugging \eqref{SADP 3}, \eqref{SBDP}, \eqref{Rate 1} and \eqref{Rate 2} into \eqref{Rate calculation}, we can get the average successful delivery rate in general scenario.

\begin{remark}
Average successful delivery rate is consist of two parts: the basic rate demand and the average of the extra rate exceeding the rate demand.
\end{remark}

Two special cases containing the interference-limited scenario and the mean load scenario are derived to further simplify the average successful delivery rate.
\begin{corollary}\label{corollary:Rate no noise}
The average successful delivery rate in the interference-limited scenario is given by
\begin{align}\label{Rate no noise}
&{\tilde R^{suc}} = {Q_{Mp}}\tilde R_{Mp,1}^{suc} + {Q_{Lp}}\tilde R_{Lp,2}^{suc} + {Q_{Lp}}{A_{Lp,1}}\tilde C_{Lp,1}^wC_{Lp,1}^b{R_0},\nonumber \\
&\tilde R_{Mp,1}^{suc} = {R_0} \nonumber \\
&+ \int_{{R_0}}^\infty  \sum\limits_{{l_1} = 0}^\infty  {\left( {{{G^ * }\left( {\delta _1^r,\delta_1 ,\alpha } \right) + {{\hat \lambda }_{2,1}}\hat P_{2,1}^{2/\alpha }{H^ * }\left( {\delta _1^r,\delta_1 ,\alpha } \right) + 1}}\right) ^{-1}} \nonumber \\
&\times {P_{{L_1}}}({l_1} + 1)dr, \nonumber \\
&\tilde R_{Lp,2}^{suc} = {A_{Lp,2}}{R_0} \nonumber \\
&+ \int_{{R_0}}^\infty  {\sum\limits_{{l_2} = 0}^\infty {\left( {{\xi _2}{G^ * }\left( {\delta _2^r,{\delta _2},\alpha } \right) + {\varsigma _2}{H^ * }\left( {\delta _2^r,{\delta _2},\alpha } \right) + {\zeta _2}} \right)}^{-1} }\nonumber \\
&\times {P_{{L_2}}}({l_2} + 1)dr,
\end{align}

where $\tilde C_{Lp,1}^w$, $C_{Lp,1}^b$ are given as \eqref{SBDP NO noise}, \eqref{SBDP}.
\end{corollary}

\begin{corollary}\label{corollary:Rate mean load}
The average successful delivery rate with the mean load in the interference-limited scenario is given by
\begin{align}\label{Rate 2}
&{{\bar R}^{suc}} = {Q_{Mp}}\bar R_{Mp,1}^{suc} + {Q_{Lp}}\bar R_{Lp,2}^{suc} + {Q_{Lp}}{A_{Lp,1}}\bar C_{Lp,1}^w\bar C_{Lp,1}^b{R_0},\nonumber \\
&\bar R_{Mp,1}^{suc} = {R_0} \nonumber \\
&+ \int_{{R_0}}^\infty  {\left( {{{G^ * }\left( {{\bar \delta _1^r},{{\bar \delta }_1} ,\alpha } \right) + {{\hat \lambda }_{2,1}}\hat P_{2,1}^{2/\alpha }{H^ * }\left( {{\bar \delta _1^r},{{\bar \delta }_1} ,\alpha } \right) + 1}}\right) ^{-1}dr},\nonumber \\
&\bar R_{Lp,2}^{suc} = {A_{Lp,2}}{R_0} \nonumber \\
&+ \int_{{R_0}}^\infty {\left( {{\xi _2}{G^ * }\left( {{\bar \delta _2^r},{{\bar \delta }_2},\alpha } \right) + {\varsigma _2}{H^ * }\left( {{\bar \delta _2^r},{{\bar \delta }_2},\alpha } \right) + {\zeta _2}} \right)}^{-1} dr,
\end{align}
where $\bar \delta _1^r = {2^{{{\bar L}_1}\frac{r}{W}}} - 1$, $\bar \delta _2^r = {2^{{{\bar L}_2}\frac{r}{W}}} - 1$, ${\bar \delta _1} = {2^{{{\bar L}_1}\frac{{{R_0}}}{W}}} - 1$, ${\bar \delta _2} = {2^{\left( {{{\bar L}_2} + 1} \right)\frac{{{R_0}}}{W}}} - 1$, $\bar C_{Lp,1}^w$, $\bar C_{Lp,1}^b$ are given as \eqref{SCDP mean load}.
\end{corollary}
The proof process of \textbf{Corollary \ref{corollary:Rate no noise}}, \textbf{Corollary \ref{corollary:Rate mean load}} are similar to the \textbf{Corollary \ref{corollary:SCDP NO noise}}, \textbf{Corollary \ref{corollary:SCDP mean load}}.
\subsection{Energy Efficiency}
Throughput is defined as the successful delivery throughput in the cache-enabled HetNets and can be expressed as
{
\begin{align}
{T_{suc}} = {\lambda _u}C{R_{suc}},
\end{align}}
where $C$ is the SCDP, ${R_{suc}} $ is the average successful delivery rate. ${{{\tilde T}^{suc}}}= {{\lambda _u}\tilde C{{\tilde R}^{suc}}}$, ${{{\bar T}^{suc}}}= {{\lambda _u}\bar C{{\bar R}^{suc}}}$ are used for the interference-limited scenario and the mean load scenario.

The total power of the cache-enabled HetNets is the power summation over the active BSs of two tiers: $Pow = \sum\nolimits_{k = 1}^2 {{\lambda _k}} Po{w_k}$. The power consumed at a BS contains the BSs power consumption, the cache power consumption and the backhaul power consumption
{
\begin{align}\label{Pow total}
Po{w_k} = Pow_k^s + Pow_k^c + Pow_k^b.
\end{align}}

BS power consumption consists of the transmit power consumption $P_k^t$ and the circuit power consumption $P_k^0$, and is expressed as
{
\begin{align}\label{BS Pow}
Pow_k^s = {\varepsilon _k}{p_{k,a}}P_k^t + P_k^0,
\end{align}}
where ${p_{k,a}}$ is the active probability of the BSs in the $k$ th tier and satisfies ${p_{1,a}} = 1$, ${p_{2,a}} = {p_a}$. ${\varepsilon _k}$ is the power amplifier efficiency.

Energy-proportional model is adopted to describe the cache power consumption. In this model, the cache power consumption of the BSs in the $k$ th tier is proportional to the cache capacity, and is expressed as
{
\begin{align}\label{cache Pow}
Pow_k^c = \rho {N_k}F,
\end{align}}
where $\rho$ is the power coefficient of cache hardware.

Backhaul power consumption depends on the successful backhaul delivery throughput. For the cases of helpers, the backhaul power consumption is zero due to the lack of the backhaul link. For the cases of MBSs, the backhaul power consumption is proportional to the successful backhaul delivery throughput calculated by multiplying the backhaul usage ratio with the backhaul capacity.
{
\begin{align}\label{backhaul Pow}
&Pow_1^b = \omega {T_b} \nonumber \\
&= \omega {C_b}\sum\limits_{{l_1} = 1}^{\infty}\sum\limits_{m = 0}^{{l_1}} {{l_1} \choose m} {\left( {1 - {p_b}} \right)^{{l_1} - m}}{\left( {{p_b}} \right)^m}\min \left\{ {1,\frac{m}{{{N_b}}}} \right\},
\end{align}}
where $\omega$ is the power coefficient of the backhaul.

With the approximation of the MBSs¡¯ mean load, backhaul power consumption can be simplified as
{
\begin{align}\label{backhaul mean Pow}
\overline {Pow_1^b}  = \omega {C_b}\sum\limits_{m = 1}^{{{\bar L}_1}} {{{\bar L}_1} \choose m} {\left( {1 - {p_b}} \right)^{{{\bar L}_1} - m}}{\left( {{p_b}} \right)^m}\min \left\{ {1,\frac{m}{{{N_b}}}} \right\},
\end{align}}
where $\bar L$ is shown as \eqref{Mean Load}.

Substituting \eqref{BS Pow}, \eqref{cache Pow} and \eqref{backhaul Pow} into \eqref{Pow total}, the total power is obtained. The total power with the mean load approximation replaces \eqref{backhaul mean Pow} with \eqref{backhaul Pow}.

According to the definition, the energy efficiency of the cache-enabled cellular networks for various cases are given in the sequel by dividing the throughput by the total power consumption. The energy efficiency in the general scenario can be derived as
\begin{align}\label{EE general}
{\eta _{EE}} = \frac{{{T^{suc}}}}{{Pow}} = \frac{{{\lambda _u}C{R^{suc}}}}{{Pow}}.
\end{align}

The energy efficiency in the interference-limited scenario can be derived as
\begin{align}\label{EE no noise}
{\tilde \eta _{EE}} = \frac{{{{\tilde T}^{suc}}}}{{Pow}} = \frac{{{\lambda _u}\tilde C{{\tilde R}^{suc}}}}{{Pow}}.
\end{align}

The energy efficiency using the mean load approximation in the interference-limited scenario can be derived as
\begin{align}\label{EE mean}
{\bar \eta _{EE}} = \frac{{{{\bar T}^{suc}}}}{{\overline {Pow} }} = \frac{{{\lambda _u}\bar C{{\bar R}^{suc}}}}{{\overline {Pow} }}.
\end{align}
\section{Numerical Results}
In this section, both the numerical simulations and Monte Carlo simulations are presented. The scenario of the two-tier HetNets is considered, in which MBSs and helpers are distributed according to PPP in a circular area with 200 km radius. The basic simulation parameters are listed in Table I and the results are averaged over 10000 Monte Carlo trials~\cite{6056691}~\cite{6364320}. We validate the theoretical analysis in the previous section via Monte Carlo simulations and compare it with the most popular caching policy in which MBSs and helpers separately select the ${N_1}$ most popular contents and the ${N_2}$ following less popular contents to store. The impact of various parameters on the system performance are investigated.
\begin{table}[h]
\vspace{-0.3em}
\centering
\captionsetup{font={scriptsize}}
\caption{SIMULATION PARAMETERS}
\begin{tabular}{|c|c|}
\hline
\textbf{Parameter} & \textbf{Value} \\
\hline
UEs density & ${\lambda _u} = 100 \times {10^{ - 6}}{\rm{ }}{m^{ - 2}}$ \\
\hline
BSs density & ${\lambda _1} = 2 \times {10^{ - 6}}{\rm{ }}{m^{ - 2}}$ \\
\hline
Bandwidth & $W = 10{\rm{ }}MHz$ \\
\hline
Path loss exponent & $\alpha  = 4$ \\
\hline
Rate threshold & ${R_0} = 100{\rm{ }}Kbps$ \\
\hline
Transmit power & $P_1^t = 46{\rm{ }}dBm$, $P_2^t = 21{\rm{ }}dBm$ \\
\hline
Static power consumption & $P_1^0 = 724.6{\rm{ }}W$, $P_2^0 = 10.16{\rm{ }}W$\\
\hline
BS power coefficient & ${\varepsilon _1} = 3.22$, ${\varepsilon _2} = 15.13$\\
\hline
Content catalog & $N = 2000$\\
\hline
Cache capacity  & $N_1 = 400$\\
\hline
Content size & $F = 10{\rm{ }}Mbit{\rm{ }}$\\
\hline
Caching power coefficient& $\rho  = 6.25 \times {10^{ - 12}}{\rm{ }}W/bit$ \\
\hline
Backhaul power coefficient & $\omega  = 5 \times {10^{ - 7}}{\rm{ }}W/bps$ \\
\hline
\end{tabular}
\vspace{-0.8em}
\end{table}

The performance of SCDP with respect to different system parameters including the helper density, content library size, helper cache capacity and backhaul capacity are shown in Fig. 2- Fig. 5. It can be seen from the figures that the simulation curves match with the numerical curves, which gives an effective validation of the theoretical analysis. The general result and the interference-limited result obtained respectively from \eqref{SCDP calculation} and \textbf{Corollary \ref{corollary:SCDP NO noise}} match. It is confirmed that noise is not a very important factor in interference-limited HetNets. The shape of the curves for the general scenario and the mean load scenario obtained from \textbf{Corollary \ref{corollary:SCDP mean load}} are consistent. The small gap under some parameter settings is caused due to the fact that the approximation error of the mean load is aggravated with the operation of summation in \eqref{SBDP}.
\begin{figure}[htbp]
\centering
\includegraphics[height=6cm,width=7.5cm]{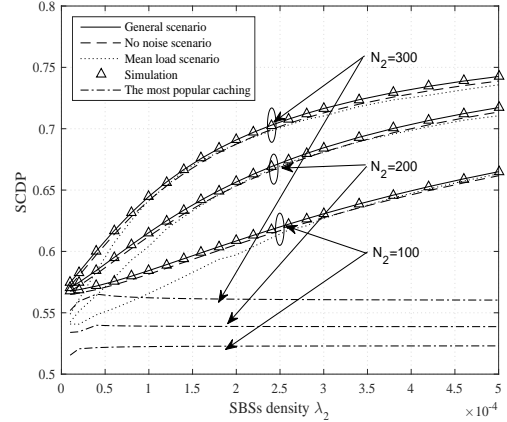}
\caption{SCDP versus helper density with different cache capacity, $\delta=0.5$, $C_b$=2.5 $\times {10^{ 6}}$ bps.}
\end{figure}

Fig. 2 shows the evolution of the SCDP with regard to the helper density for different helper cache capacity. Hybrid caching policy outperforms the most popular caching policy and the performance gain increases with a growing helper density. The reason is that numerous users are offloaded to the helpers tier and obtain the content from the local cache as the helper density increases, which is different form the most popular caching policy whose load distribution is independent of the helper density. By avoiding the influence of the limited MBS backhaul link as much as possible, dense helpers deployment can increase the delivery rate dramatically and improve the SCDP. Comparing the curves of different helper cache capacity, caching at helpers also helps to improve the SCDP.

\begin{figure}[htbp]
\centering
\includegraphics[height=6cm,width=7.5cm]{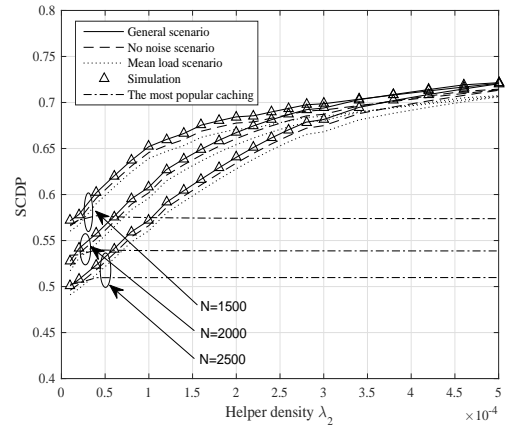}
\caption{SCDP versus helper density with different content library size, $C_b$= 2.5$\times {10^{ 6}}$ bps, $\delta=0.5$.}
\vspace{-0.4cm}
\end{figure}

Fig. 3 illustrates the evolution of SCDP with respect to the helper density for different content library size $N$. For the fixed content library size, the SCDP increases with the increase of the helper density. The performance gain of the hybrid caching policy over the most popular caching policy increases with the increase of the helper density as well. As content library size increases, the SCDP decreases. When the content library contains a large number of contents, the span of the content popularity distribution is large and the popularity of each content is relatively small given the fixed $\delta$. On the one hand, the cache hit ratio of the first content group decreases. On the other hand, the cache probability of the second content group decreases. As a result, UEs requesting content from the second group prefer to associate with the MBSs tier, the SBDP decreases accordingly. SADP $C_{Lp,2}$ decreases due to the low helper density caching the requested content. Increasing the helper density can effectively cope with the expansion of the content library to achieve the fixed SDCP requirements.
\begin{figure}[htbp]
\centering
\includegraphics[height=6cm,width=7.5cm]{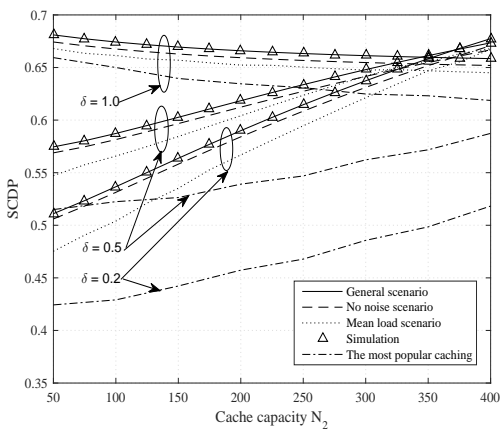}
\caption{SCDP versus helper cache capacity with different content popularity parameter, ${\lambda _2}$= 1$\times {10^{ -4}}$ m$^{-2}$, $C_b$= 2.5$\times {10^{ 6}}$ bps.}
\vspace{-0.3cm}
\end{figure}

Fig. 4 depicts the impact of the helper cache capacity on the SCDP for different content popularity distribution parameter $\delta$. The cache probability of the content in the second group increases as cache capacity increases. For a small parameter indicating the more even content popularity distribution, the SCDP gain of the proposed caching policy over the most popularity caching policy increases with the increasing helper cache capacity. For instance, setting $\delta=0.5$, the performance gain ranges from 10$\%$ to 19$\%$, when $N_2$ varies from 25 to 400. The reason for this trend is that the increasement of cache capacity helps to make full use of the content diversity to improve the SCDP. For a big parameter indicating the more skewed content popularity distribution, a large number of users¡¯ requests focus on the popular contents in the first group and MBS caching gives a large SCDP. The gap between two caching policy shrinks. However, as cache capacity increases, more users requesting the content from the second group associate with the helpers tier and active probability of the helpers increases accordingly. The SCDP experiences a slight decline under the influence of increasing inter-tier interference.
\begin{figure}[htbp]
\vspace{-0.5cm}
\centering
\includegraphics[height=6cm,width=7.5cm]{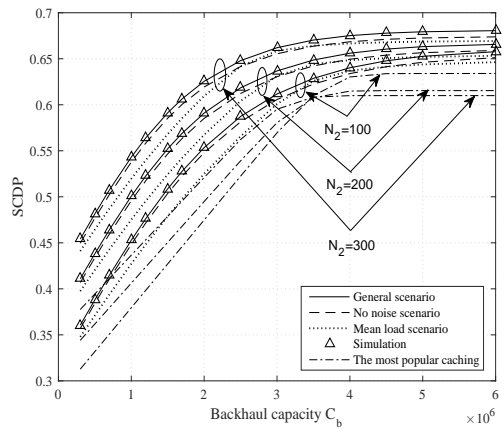}
\caption{SCDP versus backhaul capacity with different helper cache capacity, ${\lambda _2}$= 1$\times {10^{ -4}}$ m$^{-2}$, $\delta=0.5$.}
\vspace{-0.3cm}
\end{figure}

In Fig. 5, the SCDP variation with regard to backhaul capacity for different helper cache capacity is illustrated. It is shown that the increment of backhaul capacity can effectively improve the SCDP by increasing the SBDP of the UEs requesting content from the second group and associating with the MBSs tier. Each curve eventually converges to a point at which backhaul link is sufficient to assure the content availability for the cache miss UEs. Despite the increase in backhaul capacity, the gap between the hybrid caching policy and the most popular caching policy remains almost the same. As the helper cache capacity increases, the benefit of the BSs caching increase to make up for the limited backhaul link and the SCDP is improved. The larger the helper cache capacity, the more quickly the curve converges, causing the left shift of the MBSs backhaul capacity point to reach the maximum SCDP. The average performance gain increases with the increment of the cache capacity, ranging from 9.6$\%$ to 19.2$\%$ when $N_2$ varies from 100 to 300.

The performance of average successful delivery rate, energy efficiency under different system parameters containing the helper density, content library size, helper cache capacity and backhaul capacity are shown in Fig. 6- Fig. 9. It can be seen from the figures that the simulation curve is in agreement with the numerical curve. The general result and the interference-limited result obtained respectively from \eqref{Rate calculation}, \eqref{EE general} and \textbf{Corollary \ref{corollary:Rate no noise}}, \eqref{EE no noise} match. The general result and the mean load approximation obtained from \textbf{Corollary \ref{corollary:Rate mean load}}, \eqref{EE mean} are consistent. Small difference exists due to the fact that the operation of summation in \eqref{SBDP} and integral in \eqref{Rate 2} aggravate the approximation error.
\begin{figure*}[t!]
\centering
\subfigure[]
{\includegraphics[width= 0.42\linewidth, height=0.32\linewidth]{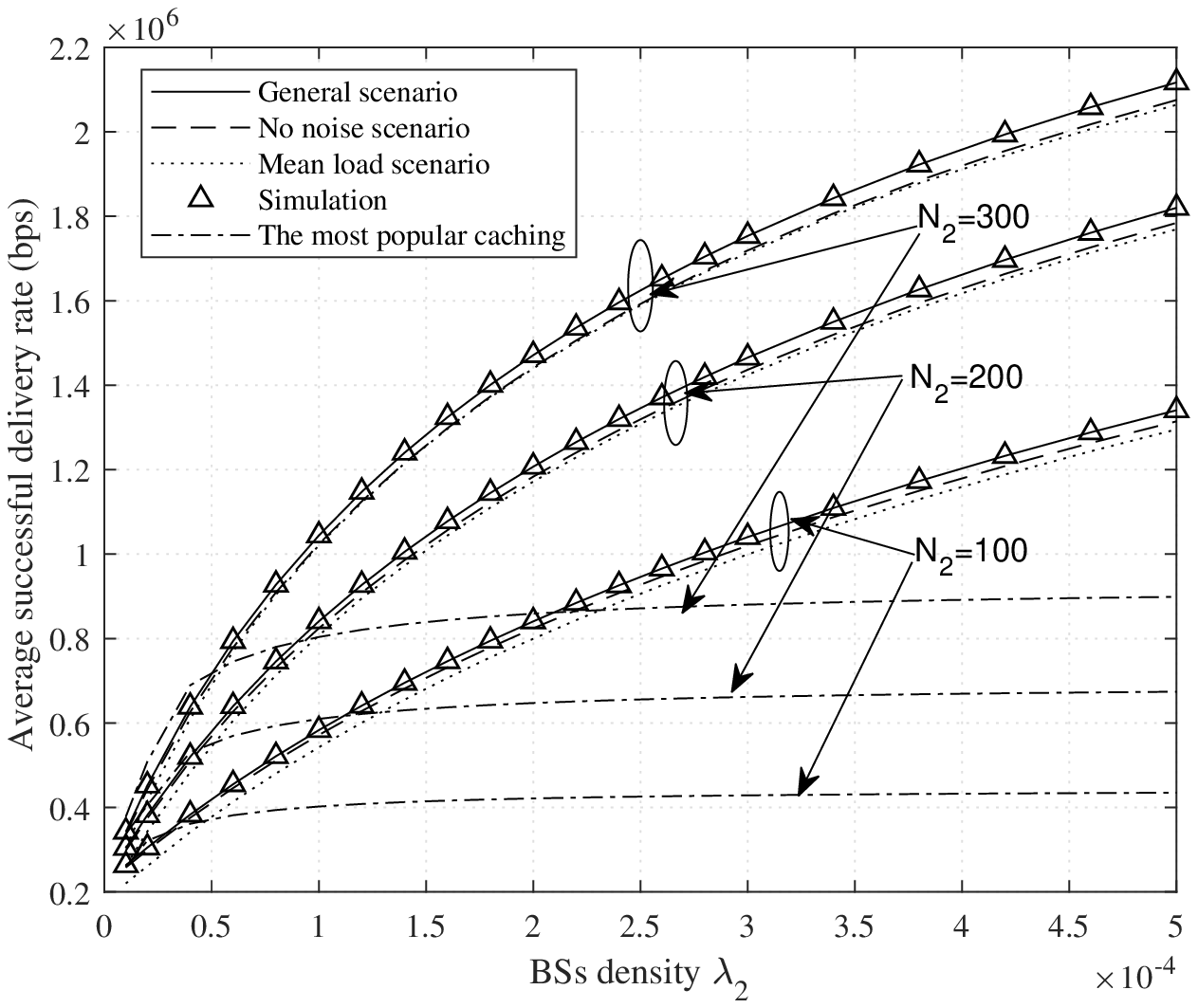}}
\subfigure[]
{\includegraphics[width= 0.42\linewidth, height=0.32\linewidth]{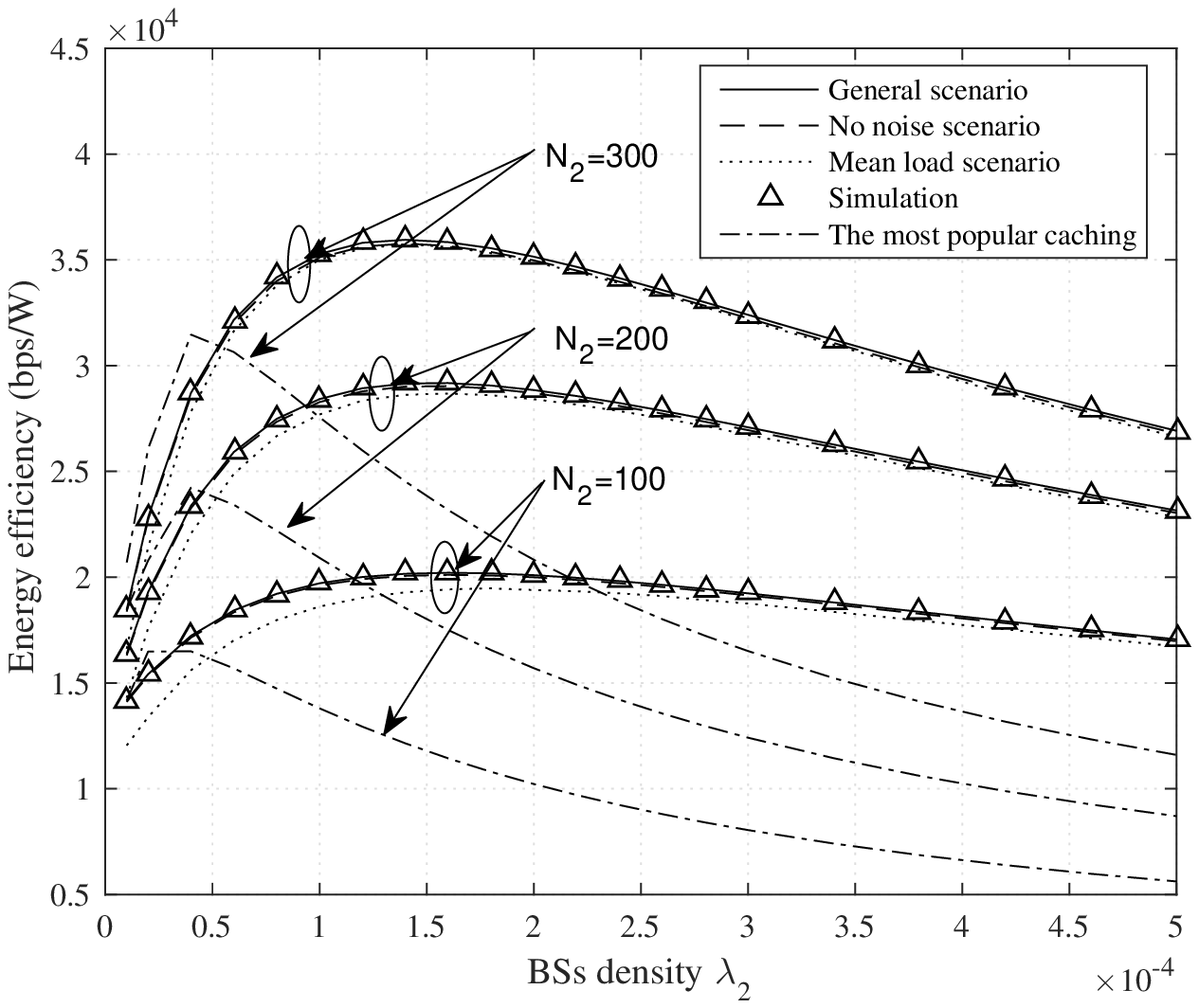}}
\caption{Average successful delivery rate (a) and energy efficiency (b) versus helper density with different helper cache capacity, $\delta=0.5$, $C_b$=2.5 $\times {10^{ 6}}$ bps.}
\vspace{-0.4cm}
\end{figure*}

\begin{figure*}[t!]
\centering
\subfigure[]
{\includegraphics[width= 0.42\linewidth, height=0.32\linewidth]{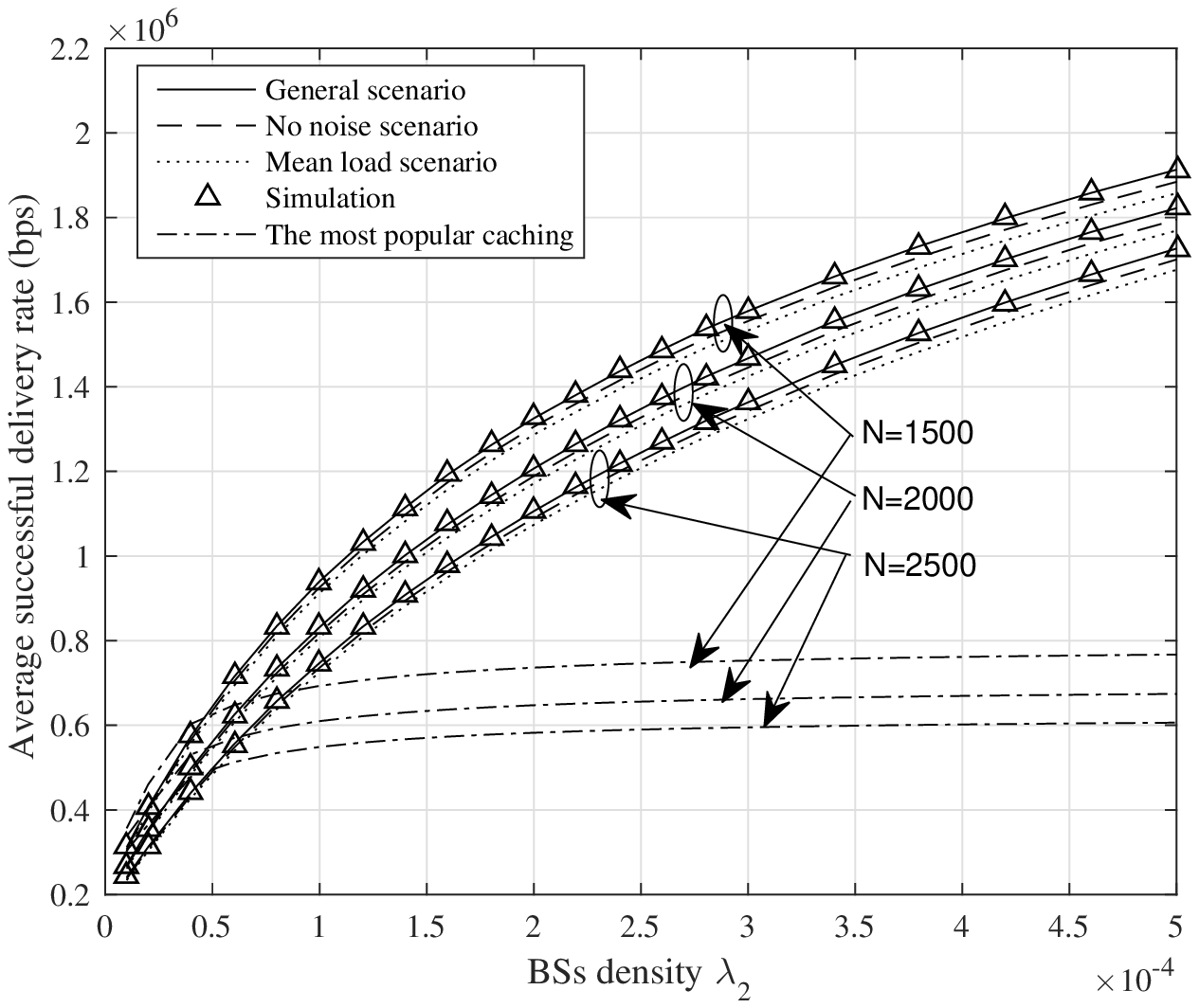}}
\subfigure[]
{\includegraphics[width= 0.42\linewidth, height=0.32\linewidth]{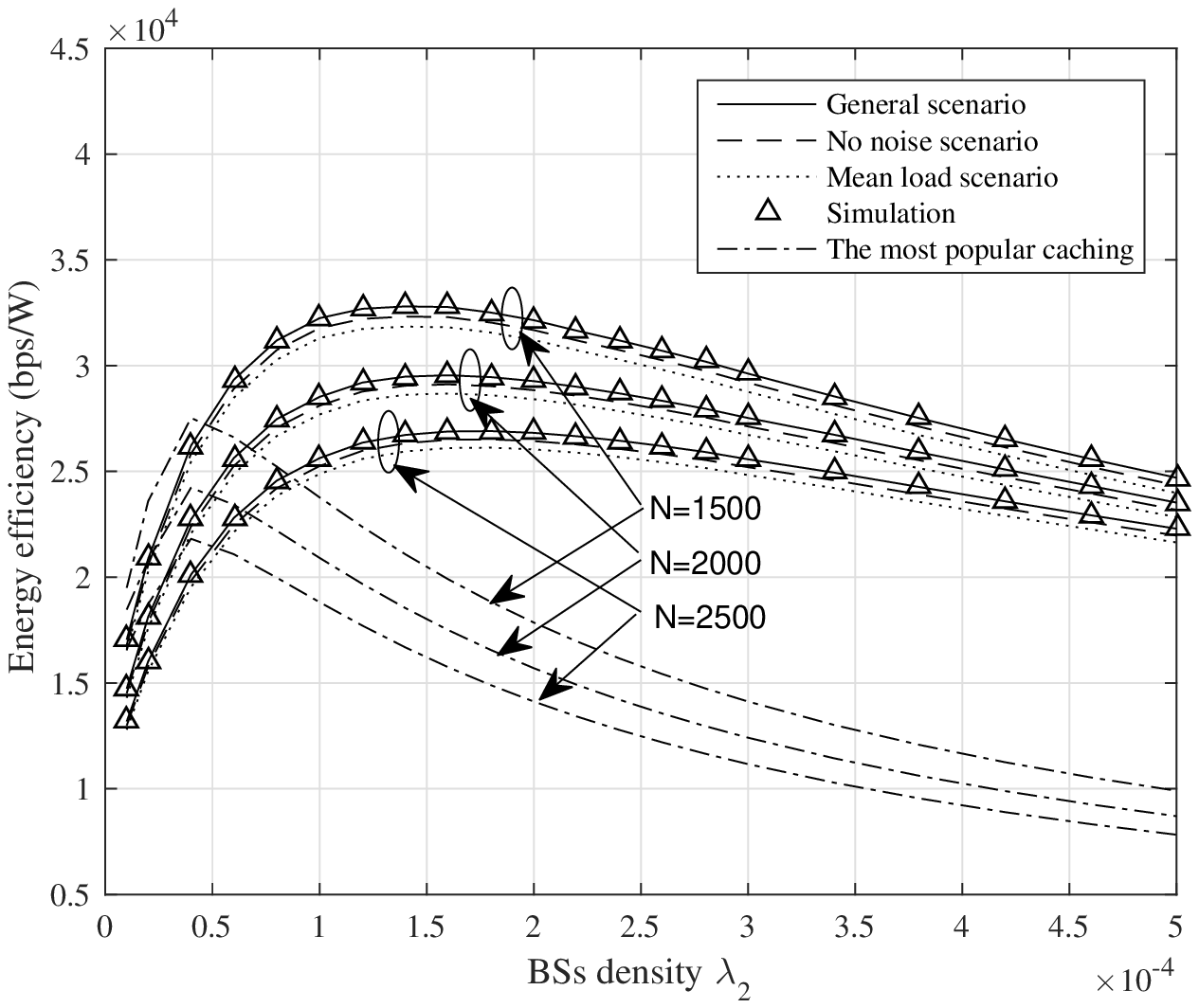}}
\caption{Average successful delivery rate (a) and energy efficiency (b) versus helper density with different content library size, $\delta=0.5$, $C_b$=2.5 $\times {10^{ 6}}$ bps.}
\vspace{-0.4cm}
\end{figure*}

The variations of the average successful delivery rate and the energy efficiency with regard to the helper density for different helper cache capacity are illustrated in Fig.6. In Fig. 6a, as helper density increases, a large number of users are offloaded to the helpers tier. Dense helpers deployment greatly increases the access delivery rate and improves the average successful delivery rate. The increment of the helper cache capacity further enhance the average successful delivery rate. In contrast, the average successful delivery rate achieved by the most popular caching policy is far less than that of the hybrid caching policy and exhibits a slight improvement as helper density increases. The reason is that the load of different BSs tier is fixed. In Fig. 6b, for small helper density, the curves of two caching policy almost coincide. As the helper density increases, the EE first increases rapidly benefitting from the improvement of the throughput, then decreases on account of the large power consumption, especially the most popular caching policy suffers a more severe degradation. There exists an optimal helper density to maximize the EE. BSs caching is able to improve the EE due to the increment of the delivery rate for the one side and the reduction of the backhaul power consumption for the other side. The larger the cache capacity, the smaller the helper density to get the maximum EE.

Fig. 7 illustrates the evolution of the average successful delivery rate and the energy efficiency with respect to the helper density for different content library size $N$.  In Fig. 7a, for the fixed content library size, the average successful delivery rate increases with the increment of the helper density. Hybrid caching policy outperforms the most popular caching policy. The average successful delivery rate decreases with the increasing content library size. As content library size increases, the cache hit ratio of the first content group decreases for the reason that MBS with a fixed cache capacity can only store small parts of the contents. Moreover, the cache probability of the second content group decreases, the probability that UEs requesting content from the second group associate with the MBSs tier increases accordingly, which result in the decrease of the average successful backhaul delivery rate. The average successful access delivery rate of the helpers decreases due to the decreasing density of the helpers capable of accessing the requested content. In Fig. 7b, as the helper density increases, the EE increases first and then decreases. EE decreases with the increasing content library size as a result of the decreasing average successful delivery rate. The larger the content library size , the bigger the helper density to get the maximum EE.

\begin{figure*}[t!]
\centering
\subfigure[]
{\includegraphics[width= 0.42\linewidth, height=0.32\linewidth]{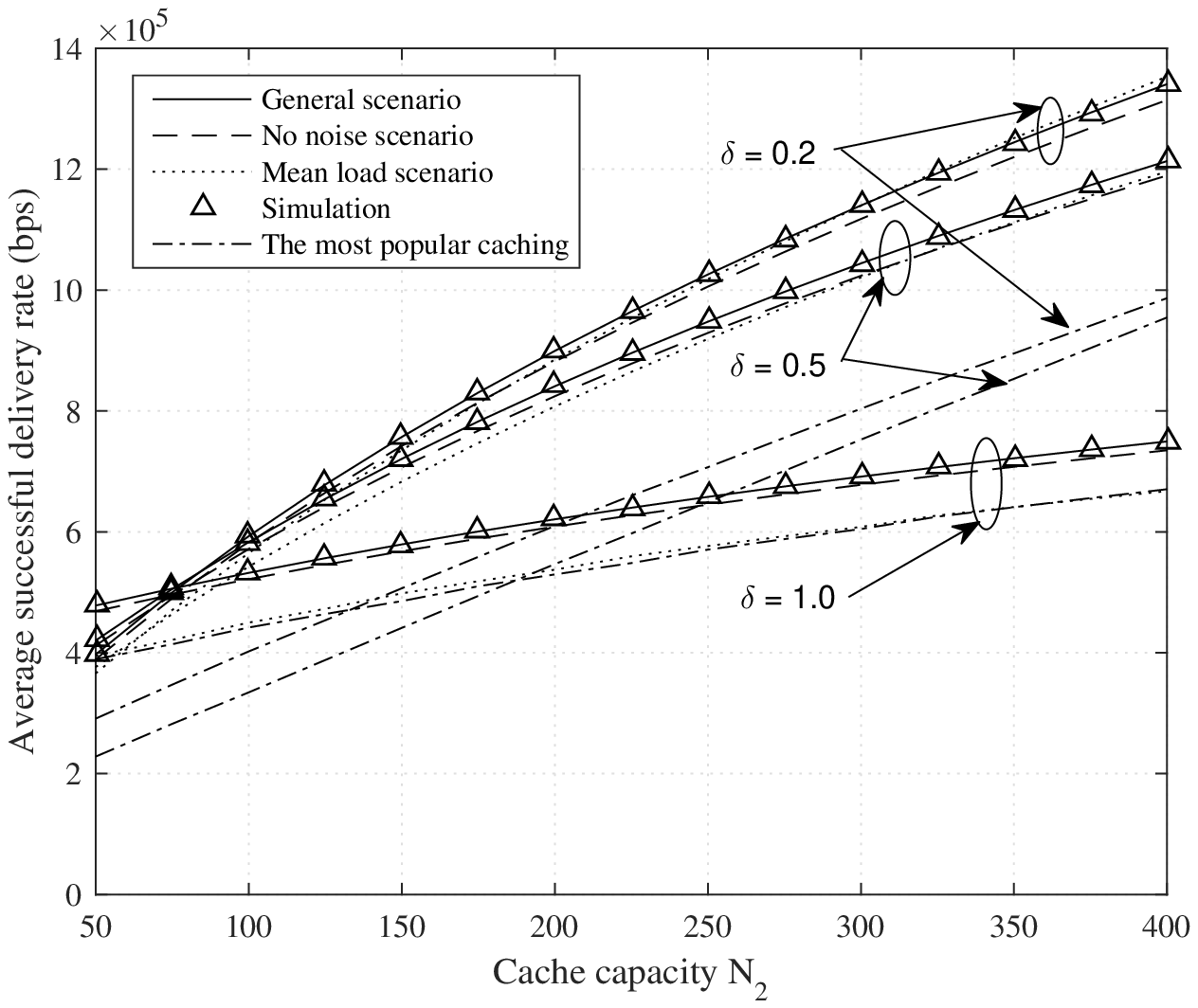}}
\subfigure[]
{\includegraphics[width= 0.42\linewidth, height=0.32\linewidth]{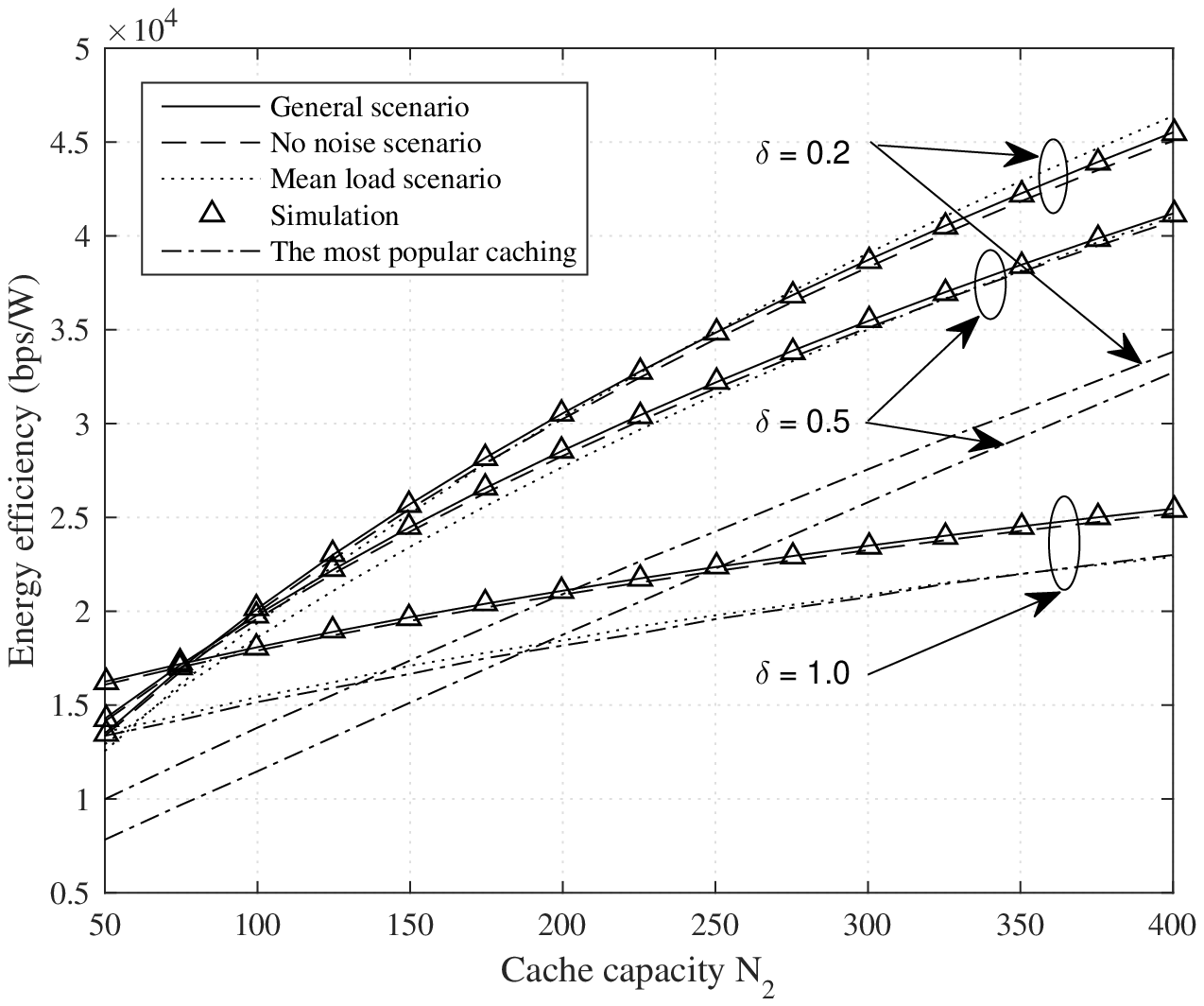}}
\caption{Average successful delivery rate (a) and energy efficiency (b) versus helper cache capacity with different content popularity parameter, ${\lambda _2}$= 1$\times {10^{ -4}}$ m$^{-2}$, $C_b$= 2.5$\times {10^{ 6}}$ bps.}
\vspace{-0.4cm}
\end{figure*}

The variations of the average successful delivery rate and the energy efficiency with regard to the helper cache capacity for different content popularity distribution parameter are illustrated in Fig. 8. In Fig. 8a, the increment of the helper cache capacity lead to dramatic improvements in the average successful delivery rate  owing to the overwhelming advantages of the access delivery rate obtained by the helper caching over the limited backhaul capacity installed in the MBSs. The hybrid caching policy achieve higher  than the most popular caching policy and the performance gain increases greatly with a growing cache capacity.  A big $\delta$ indicates that a large number of users tend to request the most popular contents in the first group and associate with the MBSs tier. The heavy load of the MBSs decreases the access delivery rate of the MBSs and results in the serious decline of the average successful delivery rate. The trend of the EE curve in Fig. 8b is the same with average successful delivery rate  curve  in Fig. 6a due to the fact that the successful delivery throughput depends on the average successful delivery rate and the cache power is quite small compared with the BSs power consumption. The EE increases with the increasing helper cache capacity. The hybrid caching policy performs better than the most popular caching policy in terms of the EE and the performance gain increases with the increasing cache capacity, the maximum gain can reach 41$\%$ for $\delta=0.2$.  As parameter $\delta$ increases, the EE decreases due to the sharp reduction of the throughput of the MBSs tier which makes up a large portion of the total network throughput. The EE gain over the most popular caching policy decreases.
\begin{figure*}[t!]
\centering
\subfigure[]
{\includegraphics[width= 0.42\linewidth, height=0.32\linewidth]{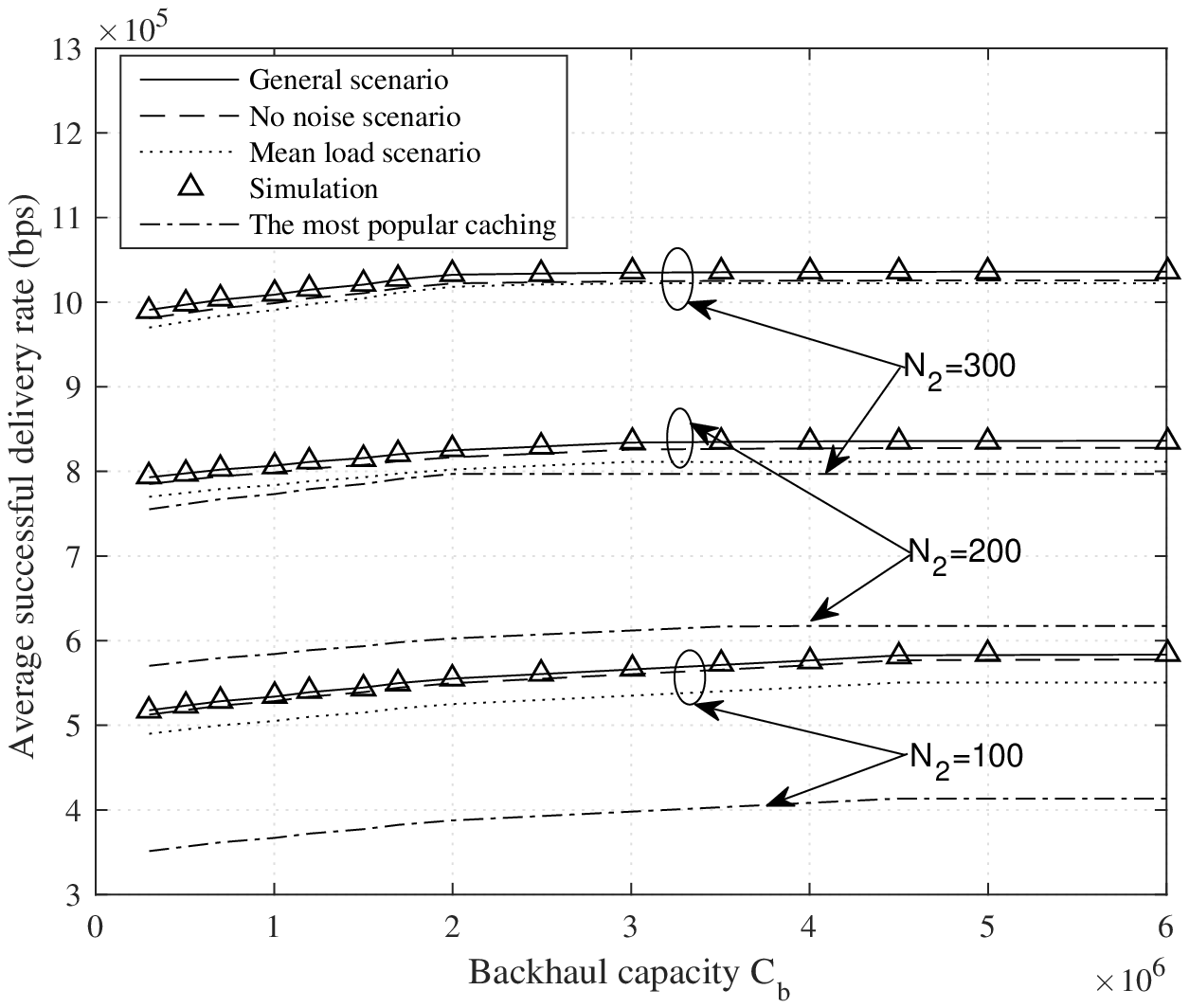}}
\subfigure[]
{\includegraphics[width= 0.42\linewidth, height=0.32\linewidth]{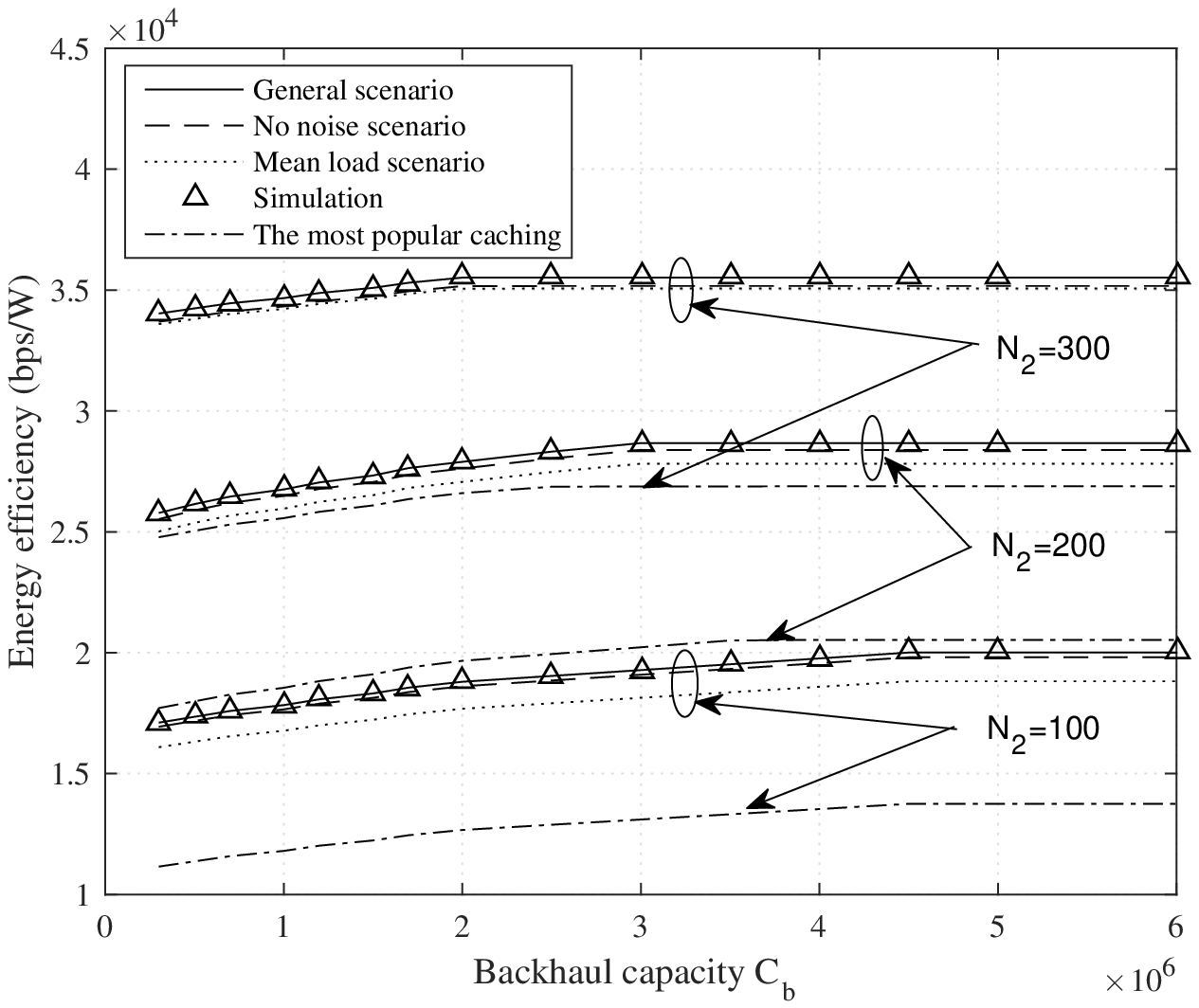}}
\caption{Average successful delivery rate (a) and energy efficiency (b) versus backhaul capacity with different helper cache capacity, ${\lambda _2}$= 1$\times {10^{ -4}}$ m$^{-2}$, $\delta=0.5$.}
\vspace{-0.4cm}
\end{figure*}

The variations of the average successful delivery rate and the energy efficiency with regard to backhaul capacity for different helper cache capacity are illustrated in Fig. 9. In Fig. 9a, the increment of the backhaul capacity increases the backhaul delivery rate of the cache miss UEs requesting the content from the second group and improves the average successful delivery rate. All curves eventually approach the fixed value, indicating that there are enough backhaul capacity to support the cache miss UEs. Moreover, caching at helpers further increases the average successful delivery rate.  The average successful delivery rate of the hybrid caching policy is  higher than that of the most popular caching policy. In Fig. 9b, the EE increases with the increasing backhaul capacity until the EE gradually tends to a fixed value. Increasing the helper cache capacity enhances the EE. The larger the cache capacity, the smaller the backhaul capacity required to achieve the maximum EE. The Hybrid caching policy exhibits a significantly better EE than the most popular caching policy and the gain increases with the increasing cache capacity, reaching  32$\%$ for $N_2=300$.
\section{Conclusion}
The cache enabled Hetnets with the limited backhaul was analyzed using the stochastic geometry. A hybrid caching policy in which the most popular contents are cached in the macro BSs tier with the deterministic caching strategy and the less popular contents are cached in the helpers tier with the probabilistic caching strategy was proposed. Taking the overall consideration of the access link, the cache and the backhaul link, the content centric association strategy is designed correspondingly. New analytical expressions of successful content delivery probability, average successful delivery rate and energy efficiency for the general scenario, the interference-limited scenario and the mean load scenario were derived. Numerical results showed that the performance can be improved greatly by the hybrid caching policy compared with the most popular caching policy. The performance gain is more obvious when the content popularity is less skewed, the cache capacity is sufficient and the helper density is relatively large. For the fixed cache capacity, there existed an optimal helper density to maximize the energy efficiency.
\numberwithin{equation}{section}
\section*{Appendix~A} \label{Appendix:A}
\renewcommand{\theequation}{A.\arabic{equation}}
\setcounter{equation}{0}

According to the definition of SCDP ${C_{Mp,1}}$, we can deduce the formula as follows
\begin{align}
&{C_{Mp,1}} = {A_{Mp,1}}{\mathbb{P}}\left( {\frac{W}{{{L_1}}}{{\log }_2}(1 + {\rm SIN}{{\rm R}_{Mp,1}}) > {R_0}} \right) \nonumber \\
&= \sum\limits_{{l_1} = 0}^\infty  {\mathbb{P}}{\left( {\frac{{{P_1}{h_{Mp,1}}Z_{Mp,1}^{ - \alpha }}}{{{I_1} + {I_2} + {\sigma ^2}}} > {2^{\left( {{l_1} + 1} \right)\frac{{{R_0}}}{W}}} - 1} \right)} {P_{{L_1}}}({l_1} + 1),
\end{align}
where ${L_1}$ is the load of the MBS. Defining ${\delta _1} = {2^{\left( {{l_1} + 1} \right)\frac{{{R_0}}}{W}}} - 1$ as the equivalent SINR threshold of the MBS tier , ${C_{Mp,1}}$ can be expressed as
\begin{align}\label{SCDP Mp 1}
&{C_{Mp,1}}\mathop  = \limits^{(a)} {{\mathbb{E}}_{{I_1},{I_2}}}\exp \left( { - \frac{{Z_{Mp,1}^{ - \alpha }\left( {{I_1} + {I_2} + {\sigma ^2}} \right){\delta _1}}}{{{P_1}}}} \right)\nonumber \\
&= \int_0^\infty  {\sum\limits_{{l_1} = 0}^\infty  {\exp \left( { - \frac{{{z^\alpha }{\sigma ^2}{\delta _1}}}{{{P_1}}}} \right){{\cal L}_{{I_1}}}\left( {\frac{{{z^\alpha }{\delta _1}}}{{{P_1}}}} \right){{\cal L}_{{I_2}}}\left( {\frac{{{z^\alpha }{\delta _1}}}{{{P_1}}}} \right)} } \nonumber \\
&\times{P_{{L_1}}}({l_1} + 1){f_{{Z_{Mp,1}}}}\left( z \right)dz,
\end{align}
where equality (a) holds due to ${h_{Mp,1}} \sim \exp \left( 1 \right)$, ${\cal L}\left( \cdot \right)$ denotes the Laplace transform, ${f_{{Z_{Mp,1}}}}\left( z \right)$ is the probability density function (PDF) of the serving distance. When the tagged user requests content from the first group and connects to the MBSs tier, the serving MBSs is the closet MBSs and the PDF of the distance between the tagged user and the closet MBSs can be expressed as
\begin{align}\label{PDF Mp 1}
{f_{{Z_{Mp,1}}}}\left( z \right) = 2\pi {\lambda _1}z{e^{ - \pi {\lambda _1}{z^2}}}.
\end{align}

The interference consists of two parts. The first part ${I_1} = \sum\limits_{{x_{1,j}} \in {\Phi _1}\backslash {x_{0}}} {{P_1}{h_{1,j}}z_{1,j}^{ - \alpha }}$ comes from all MBSs except the serving MBS located outside the circle of radius $z$. The second part ${I_2} = \sum\limits_{{x_{2,j}} \in {\Phi _2}} {{P_{2}}{h_{2,j}}z_{2,j}^{ - \alpha }}$  comes from all helpers in the active mode dispersed across the entire area.

The Laplace transform of the first part of interference is derived as follows
\begin{align}
&{{\cal L}_{{I_1}}}\left( s \right) = {{\mathbb{E}}_{{I_1}}}\left( {{e^{ - s{I_1}}}} \right)\nonumber \\
&= {{\mathbb{E}}_{{\Phi _1},{h_{1,j}}}}\left( {\prod\limits_{{x_{1,j}}  \in {\Phi _1}\backslash {x_{0}}} {\exp \left( { - s{P_1}{h_{1,j}}z_{1,j}^{ - \alpha }} \right)} } \right)\nonumber \\
&\mathop  = \limits^{(b)} {{\mathbb{E}}_{{\Phi _1},{h_{1,j}}}}\left( {\prod\limits_{{x_{1,j}} \in {\Phi _1}\backslash {x_{0}}} {\frac{1}{{1 + s{P_1}z_{1,j}^{ - \alpha }}}} } \right) \nonumber \\
&\mathop  = \limits^{(c)} \exp \left( { - 2\pi {\lambda _1}\int_z^\infty  {\left( {1 - \frac{1}{{1 + s{P_1}y^{ - \alpha }}}} \right)ydy} } \right),
\end{align}
where (b) follows from ${h_{k,i}} \sim \exp \left( 1 \right)$, (c) follows from the probability generating function of the PPP.

Let $s = \frac{{{z^\alpha }{\delta _1}}}{{{P_1}}}$, ${L_{{I_1}}}$ is given as
\begin{align}\label{Laplace Mp 1}
{{\cal L}_{{I_1}}}\left( {\frac{{{z^\alpha }{\delta _1}}}{{{P_1}}}} \right) = \exp \left( { - \pi {\lambda _1}G\left( {{\delta _1},\alpha } \right){z^2}} \right),
\end{align}
where $G\left( {{\delta _1},\alpha } \right) = \frac{{2{\delta _1}}}{{\alpha  - 2}}{}_2{F_1}\left( {1,1 - \frac{2}{\alpha };2 - \frac{2}{\alpha }; - {\delta _1}} \right)$.

The Laplace transform of the second part of interference is derived as follows
\begin{align}
&{{\cal L}_{{I_2}}}\left( s \right) = {{\mathbb{E}}_{{I_2}}}\left( {{e^{ - s{I_2}}}} \right)= {{\mathbb{E}}_{{\Phi _2},{h_{2,j}}}}\left( {\prod\limits_{x_{2,j} \in {\Phi _{2,a}}} {\exp \left( { - s{P_2}{h_{2,j}}z_{2,j}^{ - \alpha }} \right)} } \right)\nonumber \\
&= {{\mathbb{E}}_{{\Phi _{2,a}}}}\left( {\prod\limits_{i \in {\Phi _{2,a}}} {\frac{1}{{1 + s{P_2}z_{2,j}^{ - \alpha }}}} } \right)\nonumber \\
&=\exp \left( { - 2\pi {p_a}{\lambda _2}\int_z^\infty  {\left( {1 - \frac{1}{{1 + s{P_2}{y^{ - \alpha }}}}} \right)ydy} } \right)
\end{align}

Let $s = \frac{{{z^\alpha }{\delta _1}}}{{{P_1}}}$, ${L_{{I_2}}}$ is given as
\begin{align}\label{Laplace Mp 2}
{{\cal L}_{{I_2}}}\left( {\frac{{{z^\alpha }{\delta _1}}}{{{P_1}}}} \right) = \exp \left( { - \pi {p_a}{\lambda _2}\hat P_{2,1}^{2/\alpha }H\left( {{\delta _1},\alpha } \right){z^2}} \right),
\end{align}
where $H\left( {{\delta _1},\alpha } \right) = \frac{2}{\alpha }\delta _1^{^{2/\alpha }}B\left( {\frac{2}{\alpha },1 - \frac{2}{\alpha }} \right)$.

Integrating \eqref{Laplace Mp 1}, \eqref{Laplace Mp 2} and \eqref{PDF Mp 1} into \eqref{SCDP Mp 1} completes the proof.
\section*{Appendix~B} \label{Appendix:B}
\renewcommand{\theequation}{B.\arabic{equation}}
\setcounter{equation}{0}

According to the definition of SCDP ${C_{Lp,2}}$, we can deduce the formula as follows
\begin{align}
&{C_{Lp,2}} = {A_{Lp,2}}{\mathbb{P}}\left( {\frac{W}{{{L_2}}}{{\log }_2}(1 +{\rm SIN}{{\rm R}_{Lp,2}}) > {R_0}} \right)\nonumber \\
&= {A_{Lp,2}}\sum\limits_{{l_2} = 0}^\infty  {{\mathbb{P}}\left( {\frac{{{P_2}{h_{Lp,2}}Z_{Lp,2}^{ - \alpha }}}{{{I_1} + {L_{{I_{L{p^ + },2}}}} + {L_{{I_{L{p^ - },2}}}} + {\sigma ^2}}} > {2^{\left( {{l_2} + 1} \right)\frac{{{R_0}}}{W}}} - 1} \right)} \nonumber \\
&\times{P_{{L_2}}}({l_2} + 1)
\end{align}
where ${L_2}$ is the load of the helper. Defining ${\delta _2} = {2^{\left( {{l_2} + 1} \right)\frac{{{R_0}}}{W}}} - 1$ as the equivalent SINR threshold of the helper tier, ${C_{Lp,2}}$ can be expressed as
\small{\begin{align}\label{SCDP Lp 2}
 &{C_{Lp,2}}= {{\mathbb{E}}_{{I_1},{I_{L{p^ + },2}},{I_{L{p^ - },2}}}}\exp \left( { - \frac{{Z_{Lp,2}^{ - \alpha }\left( {{I_1} + {I_{L{p^ + },2}} + {I_{L{p^ - },2}} + {\sigma ^2}} \right){\delta _2}}}{{{P_2}}}} \right)\nonumber \\
& = {A_{Lp,2}}\int_0^\infty  \sum\limits_{{l_2} = 0}^\infty  \exp \left( { - \frac{{{z^\alpha }{\sigma ^2}{\delta _2}}}{{{P_2}}}} \right){{\cal L}_{{I_1}}}\left( {\frac{{{z^\alpha }{\delta _2}}}{{{P_2}}}} \right){{\cal L}_{{I_{L{p^ + },2}}}}\left( {\frac{{{z^\alpha }{\delta _2}}}{{{P_2}}}} \right)\nonumber \\
&\times{{\cal L}_{{I_{L{p^ - },2}}}}\left( {\frac{{{z^\alpha }{\delta _2}}}{{{P_2}}}} \right) {P_{{L_2}}}({l_2} + 1) {f_{{Z_{Lp,2}}}}\left( z \right)dz,
\end{align}}
\normalsize
where ${f_{{Z_{Lp,2}}}}\left( z \right)$ is the PDF of the serving distance when the tagged user requests content from the second group and connects to the helpers tier. Similar to the derivation of the access probability, the serving PDF can be obtained by matching the content-centric HetNets with the traditional HetNets and straightforwardly modifying the Lemmas 4 in ~\cite{6287527} as
\begin{align}\label{PDF Lp 2}
{f_{{Z_{Lp,2}}}}\left( z \right) = \frac{{2\pi {p_{Lp}}{\lambda _2}z}}{{{A_{Lp,2}}}}\exp \left( { - \pi {\lambda _1}P_{1,2}^{2/\alpha }{z^2} - \pi {p_{Lp}}{\lambda _2}{z^2}} \right)
\end{align}

The interference consists of three parts. The first part ${I_1} = \sum\limits_{x_{1,j} \in {\Phi _1}} {{P_1}{h_{1,j}}z_{1,j}^{ - \alpha }}$ comes from all MBSs and the distance from the tagged user is larger than $\hat P_{1,2}^{1/\alpha }z$. The second part ${I_{L{p^ + },2}} = \sum\limits_{x_{2,j} \in {\Phi _{L{p^ + },2,a}}\backslash {x_{2,0}}} {{P_2}{h_{2,j}}z_{2,j}^{ - \alpha }}$ comes from the helpers storing the requested content in the active mode except the serving helper and the distance from the tagged user is larger than $z$ . The third part ${I_{L{p^ - },2}} = \sum\limits_{x_{2,j}\in {\Phi _{L{p^ - },2,a}}} {{P_2}{h_{2,j}}z_{2,j}^{ - \alpha }}$ comes from the helpers not storing the requested content in the active mode, which are dispersed across the entire area.

Taking the Laplace transform of the three parts of the interference, and integrating \eqref{PDF Lp 2} into \eqref{SCDP Lp 2}, the proof is completed.
\section*{Appendix~C} \label{Appendix:C}
\renewcommand{\theequation}{C.\arabic{equation}}
\setcounter{equation}{0}
Similar to the derivation in Appendix B, we can deduce $C_{Lp,1}^w$ as follows
\begin{align}\label{SCDP Lp 3}
&C_{Lp,1}^w ={A_{Lp,1}}\int_0^\infty  \sum\limits_{{l_1} = 0}^\infty  \exp \left( { - \frac{{{z^\alpha }{\sigma ^2}{\delta _1}}}{{{P_1}}}} \right){{\cal L}_{{I_1}}}\left( {\frac{{{z^\alpha }{\delta _1}}}{{{P_1}}}} \right)\nonumber \\
&\times{{\cal L}_{{I_{L{p^ + },2}}}}\left( {\frac{{{z^\alpha }{\delta _1}}}{{{P_1}}}} \right){{\cal L}_{{I_{L{p^ - },2}}}}\left( {\frac{{{z^\alpha }{\delta _1}}}{{{P_1}}}} \right){P_{{L_1}}}({l_1} + 1) {f_{{Z_{Lp,1}}}}\left( z \right)dz
\end{align}
where ${f_{{Z_{Lp,1}}}}\left( z \right)$ is the PDF of the serving distance when the tagged user requests content from the second group and connects to the MBSs tier. ${f_{{Z_{Lp,1}}}}\left( z \right)$ can be expressed as
\begin{align}\label{PDF Lp 3}
{f_{{Z_{Lp,1}}}}\left( z \right) = \frac{{2\pi {\lambda _1}z}}{{{A_{Lp,1}}}}\exp \left( { - \pi {p_{Lp}}{\lambda _2}\hat P_{2,1}^{2/\alpha }{z^2} - \pi {\lambda _1}{z^2}} \right)
\end{align}

The interference consists of three parts. The first part ${I_1} = \sum\limits_{x_{1,j} \in {\Phi _1}\backslash {x_{0}}} {{P_1}{h_{1,j}}z_{1,j}^{ - \alpha }}$ comes from all MBSs except the serving MBS and the distance from the tagged user is larger than $z$. The second part ${I_2} = \sum\limits_{x_{2,j}\in {\Phi _{L{p^ + },2,a}}} {{P_2}{h_{2,j}}z_{2,j}^{ - \alpha }}$ comes from the helpers storing the requested content in the active mode and the distance from the tagged user is larger than $\hat P_{1,2}^{1/\alpha }z$. The third part ${I_{L{p^ - },2}} = \sum\limits_{x_{2,j}\in {\Phi _{L{p^ - },2,a}}} {{P_2}{h_{2,j}}z_{2,j}^{ - \alpha }}$ comes from the helpers not storing the requested content in the active mode, which are dispersed across the entire area.

Taking the Laplace transform of the three parts of the interference, and integrating \eqref{PDF Lp 3} into \eqref{SCDP Lp 3}, the proof is completed.
\section*{Appendix~D} \label{Appendix:D}
\renewcommand{\theequation}{D.\arabic{equation}}
\setcounter{equation}{0}
According to the definition, $R_{Mp,1}^{suc}$ is derived as follows
\begin{align}
&R_{Mp,1}^{suc} = {A_{Mp,1}}{\mathbb{E}}\left( {{R_{Mp,1}}|{R_{Mp,1}} > {R_0}} \right)\nonumber \\
&= \int_0^\infty  {{\mathbb{P}}\left( {\frac{W}{{{L_1}}}{{\log }_2}(1 + {\rm SIN}{{\rm R}_{Mp,1}}) > r|{R_{Mp,1}} > {R_0}} \right)dr} \nonumber \\
 & = {R_0} + \int_{{R_0}}^\infty  \int_0^\infty  {\sum\limits_{{l_1} = 0}^\infty  {\frac{{{\mathbb{P}}\left( {\frac{{{P_1}{h_{Mp,1}}{z^{ - \alpha }}}}{{{I_1} + {I_2} + {\sigma ^2}}} > {2^{\left( {{l_1} + 1} \right)\frac{r}{W}}} - 1} \right)}}{{{\mathbb{P}}\left( {\frac{{{P_1}{h_{Mp,1}}{z^{ - \alpha }}}}{{{I_1} + {I_2} + {\sigma ^2}}} > {2^{\left( {{l_1} + 1} \right)\frac{{{R_0}}}{W}}} - 1} \right)}}} }\nonumber \\
 &\times{P_{{L_1}}}({l_1} + 1){f_{{Z_{Mp,1}}}}\left( z \right)dzdr
\end{align}
Let $\delta _1^r = {2^{\left( {{l_1} + 1} \right)\frac{r}{W}}} - 1$, ${\delta _1} = {2^{\left( {{l_1} + 1} \right)\frac{{{R_0}}}{W}}} - 1$
\small{
\begin{align}
&R_{Mp,1}^{suc} \mathop = \limits^{(d)} {R_0} +
\int_{{R_0}}^\infty  {\int_0^\infty  {\sum\limits_{{l_1} = 0}^\infty  {\frac{{\exp \left( { - \frac{{\delta _1^r{\sigma ^2}{z^\alpha }}}{{{P_1}}}} \right){{\mathbb{E}}_{{I_1}}}\left( {\exp \left( {\frac{{ - \delta _1^r{z^\alpha }{I_1}}}{{{P_1}}}} \right)} \right)}}{{\exp \left( { - \frac{{{\delta _1}{\sigma ^2}{z^\alpha }}}{{{P_1}}}} \right){{\mathbb{E}}_{{I_1}}}\left( {\exp \left( {\frac{{ - {\delta _1}{z^\alpha }{I_1}}}{{{P_1}}}} \right)} \right)}}} } } \nonumber\\
& \times
{\frac{{{{\mathbb{E}}_{{I_2}}}\left( {\exp \left( {\frac{{ - \delta _1^r{z^\alpha }{I_2}}}{{{P_1}}}} \right)} \right)}}
{{{{\mathbb{E}}_{{I_2}}}\left( {\exp \left( {\frac{{ - {\delta _1}{z^\alpha }{I_2}}}{{{P_1}}}} \right)} \right)}}}
 {P_{{L_1}}}({l_1} + 1){f_{{Z_{Mp,1}}}}\left( z \right)dzdr \nonumber\\
 & \mathop = \limits^{(e)} {R_0} + \int_{{R_0}}^\infty  \int_0^\infty  \sum\limits_{{l_1} = 0}^\infty  2\pi {\lambda _1}z\exp \left( { - \frac{{\left( {\delta _1^r - {\delta _1}} \right){\sigma ^2}{z^\alpha }}}{{{P_1}}}} \right)\nonumber\\
& \times {\exp \left( { - \pi {\lambda _1}{z^2}\left( {{G^ *  }\left( {\delta _1^r,{\delta _1},\alpha } \right) + {{\hat \lambda }_{2,1}}\hat P_{2,1}^{2/\alpha }{H^ * }\left( {\delta _1^r,{\delta _1},\alpha } \right) + 1} \right)} \right)} \nonumber\\
&\times {P_{{L_1}}}({l_1} + 1)dzdr
\end{align}}
\normalsize
where ${G^ * }\left( {\delta _1^r,{\delta _1},\alpha } \right) = G\left( {\delta _1^r,\alpha } \right) - G\left( {{\delta _1},\alpha } \right)$, ${H^ * }\left( {\delta _1^r,{\delta _1},\alpha } \right) = H\left( {\delta _1^r,\alpha } \right) - H\left( {{\delta _1},\alpha } \right)$, (d) follows from ${h_{Mp,1}} \sim \exp \left( 1 \right)$, (e) follows from the Laplace transforms of the interference.

\bibliographystyle{IEEEtran}
\bibliography{fanbib}

\vspace{-1.5cm}
\begin{IEEEbiography}[{\includegraphics[width=1in,height=1.25in,clip,keepaspectratio]{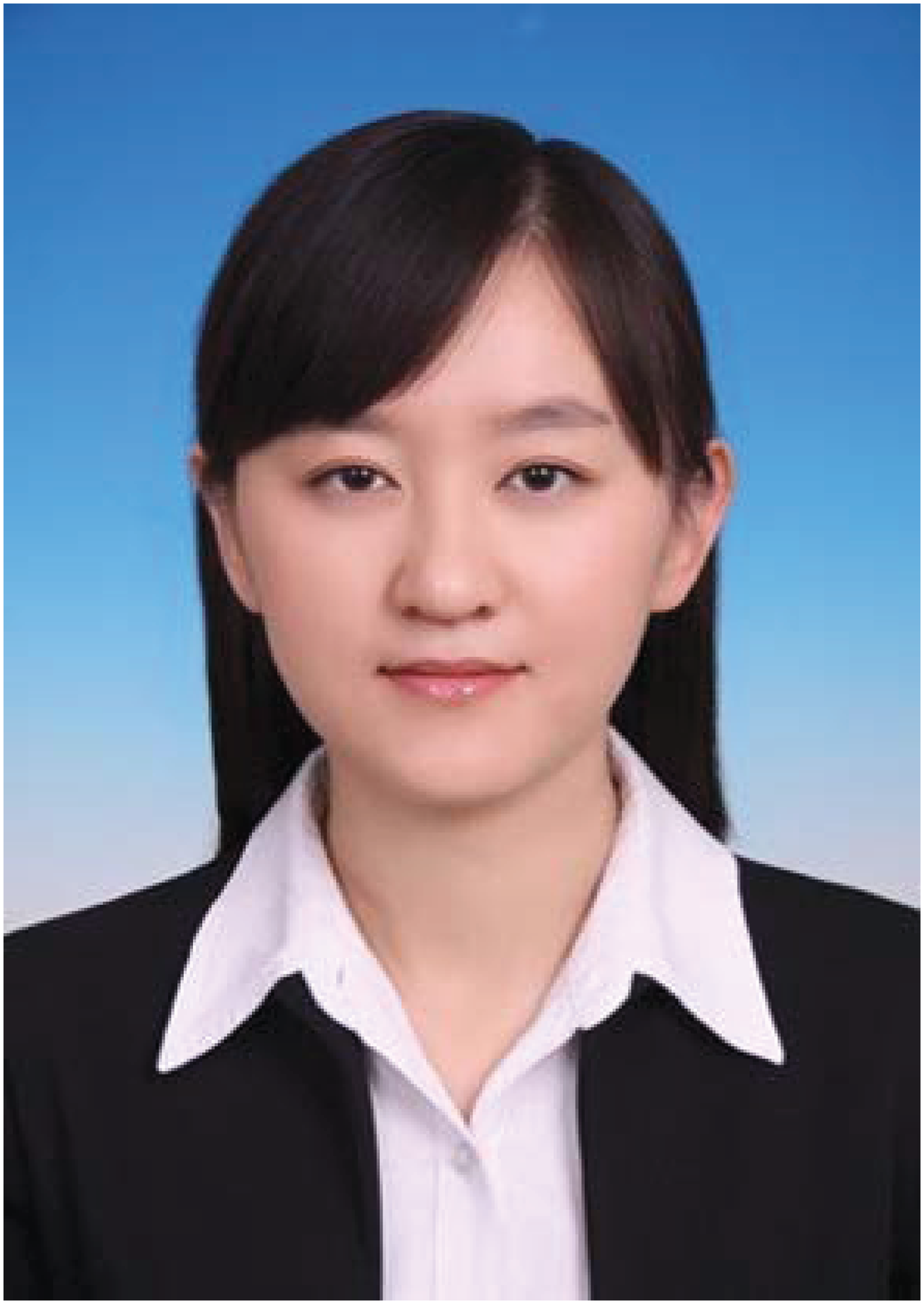}}]{Congshan Fan}
received the Ph.D. degree in Information and Communication Engineering and M.S. degree in Electronic and Communication Engineering from Beijing University of Posts and Telecommunications (BUPT), China, in 2013 and 2019, respectively.  Her current research interests include edge caching, UAVs network, ultra-dense networks.
\end{IEEEbiography}
\vspace{-1.5cm}

\begin{IEEEbiography}[{\includegraphics[width=1in,height=1.25in,clip,keepaspectratio]{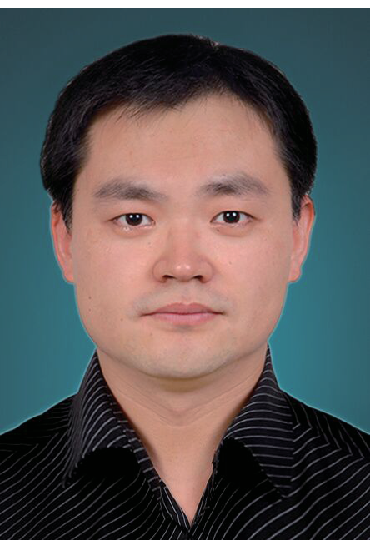}}]{Tiankui Zhang}
 (M'10-SM'15) received the Ph.D. degree in Information and Communication Engineering and B.S. degree in Communication Engineering from Beijing University of Posts and Telecommunications (BUPT), China, in 2008 and 2003, respectively. Currently, he is a Professor in School of Information and Communication Engineering at BUPT. His research interests include wireless communication networks, mobile edge computing and caching, signal processing for wireless communications, content centric wireless networks. He had published more than 100 papers including journal papers on IEEE Journal on Selected Areas in Communications, IEEE Transaction on Communications, etc., and conference papers, such as IEEE GLOBECOM and IEEE ICC.
\end{IEEEbiography}
\vspace{-1.5cm}

\begin{IEEEbiography}[{\includegraphics[width=1in,height=1.25in,clip,keepaspectratio]{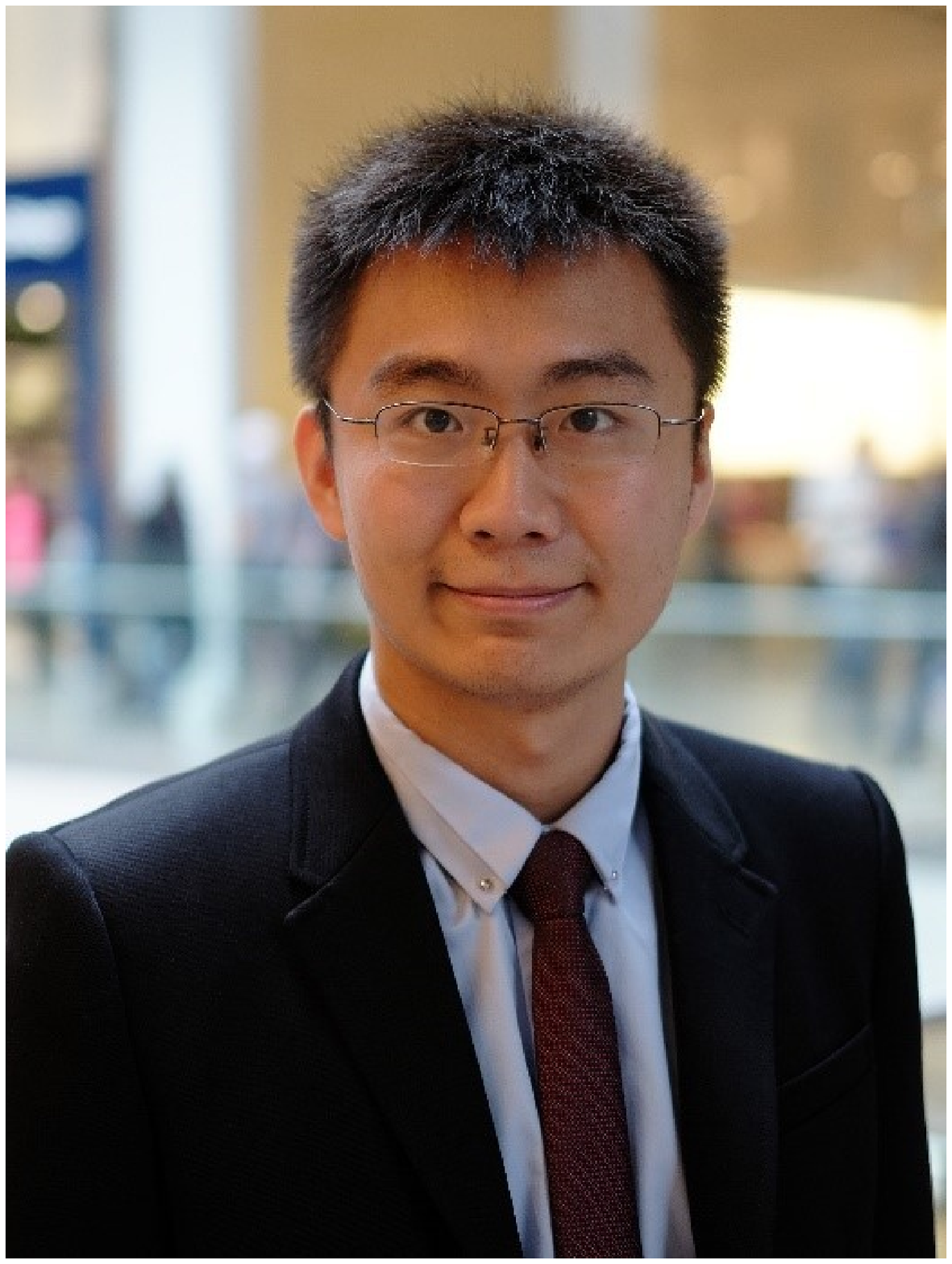}}] {Yuanwei Liu} (S'13-M'16-SM'19) received the B.S.  and M.S. degrees from the Beijing University of  Posts and Telecommunications in 2011 and 2014,  respectively, and the Ph.D. degree in electrical engineering from the Queen Mary University of London,  U.K., in 2016. He was with the Department of Informatics, King's  College London, from 2016 to 2017, where he  was a Post-Doctoral Research Fellow. He has been  a Lecturer (Assistant Professor) with the School  of Electronic Engineering and Computer Science, Queen Mary University of London, since 2017.

His research interests include  5G and beyond wireless networks, the Internet of Things, machine learning, and stochastic geometry. He has served as a TPC Member for many IEEE  conferences, such as GLOBECOM and ICC. He received the Exemplary Reviewer Certificate of IEEE WIRELESS COMMUNICATIONS LETTERS in  2015, IEEE TRANSACTIONS ON COMMUNICATIONS in 2016 and 2017, and IEEE TRANSACTIONS ON WIRELESS COMMUNICATIONS in 2017 and  2018. He has served as the Publicity Co-Chair for VTC 2019-Fall. He is currently an Editor on the Editorial Board of the IEEE TRANSACTIONS  ON COMMUNICATIONS, IEEE COMMUNICATIONS LETTERS, and IEEE ACCESS. He also serves as a Guest Editor for IEEE JSTSP special issue on Signal Processing Advances for Non-Orthogonal Multiple Access in Next Generation Wireless Networks.
\end{IEEEbiography}
\vspace{-1.5cm}

\begin{IEEEbiography}[{\includegraphics[width=1in,height=1.25in,clip,keepaspectratio]{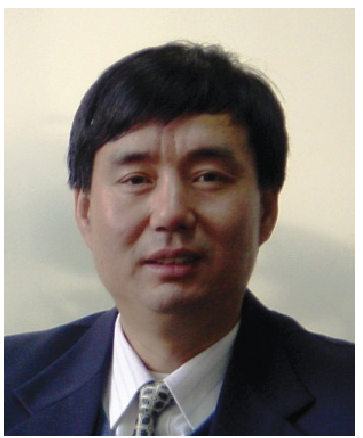}}]{Zhiming Zeng}
received the B.S. degree in carrier communication, the M.S. degree in communication and electronic systems, and the Ph.D. degree in communication and information systems from Beijing University of Posts and Telecommunication, Beijing, China. He is currently a Professor with the School of Information and Communication Engineering, Beijing University of Posts and Telecommunications. He is a Senior Member of the China Institute of Communications, an Advanced Member of the Chinese Institute of Electronics, and a member of the Academic Committee, BUPT. His current research interests include theory and technology of next generation mobile and wireless networks.
\end{IEEEbiography}
\vspace{-1.5cm}

\end{document}